**Assessing transport resilience to heavy rainfall: An integrated modelling framework with metrics**

Wei Bi [a,c,*], David Alvarez Castro [b,d], Yimeng Liu [b], Abdullah Kahraman [b,e], Alistair Ford [b], Roberto Palacin [b], Kristen MacAskill [a]

[a] Department of Engineering, University of Cambridge, Cambridge CB2 1PZ, United Kingdom
[b] School of Engineering, Newcastle University, Newcastle upon Tyne NE1 7RU, United Kingdom
[c] Institute of Construction and Infrastructure Management, ETH Zürich, 8093 Zürich, Switzerland
[d] Transport Planning Team, Arup, 28014 Madrid, Spain
[e] Tyndall Centre for Climate Change Research, Newcastle University, Newcastle upon Tyne NE1 7RU, United Kingdom

* Corresponding author.
Email address: wb316@cam.ac.uk (W. Bi) and bi@ibi.baug.ethz.ch (W. Bi).

**Abstract**

Road passenger transport faces acute disruption risks from flash flooding from heavy rainfall. To capture system complexity and dynamics for assessing its flood resilience, this study (a) proposes performance metrics to quantify disruption impacts from physical, operational, environmental, social, and economic dimensions, and (b) develops an integrated modelling framework to assess selected metrics by combining rainfall modelling, surface water flood modelling, and agent-based transport modelling to simulate asset-level disruptions and user behavioural responses. This methodology is demonstrated through a 1-in-30-year rainfall event in Tyne and Wear, North East England. Three behavioural response scenarios to rainfall are simulated, including a worst-case scenario, an ideal-response scenario, and an early-warning scenario. Results highlight the potential of behavioural adaptation in enhancing road flood resilience by reducing journey delays, cancellations, and economic losses. The framework and multidimensional performance metrics offer a transferable approach for stress-testing rainfall scenarios and evaluating interventions, supporting evidence-based climate-resilient transport planning.

## 1. Introduction

Serving as the backbone of modern societies, transport infrastructure worldwide is exposed to growing flood-related threats from extreme weather conditions, which are increasing under climate change (Koks et al., 2019; Liu et al., 2023). Particularly, short-duration heavy rainfall events often result in significant surface water flooding on road networks, overwhelming drainage systems and disrupting mobility (National Infrastructure Commission, 2022). Given the complexity of road transport as a socio-technical system, such acute rainfall-induced asset-level disruptions can quickly cascade into broader systemic impacts, leading to additional social, environmental, and economic consequences. Under these circumstances, enhancing system resilience, defined as the ability of a system to withstand and rapidly recover from adverse conditions and disruptions (Department for Transport, 2014), has become a pressing priority for transport authorities. Despite varying approaches to assessing transport resilience (Bi et al., 2023), stress testing has emerged as one of the prominent methods that evaluates system performance loss during disruption periods as a measure of system resilience (Linkov et al., 2022), based on the widely applied resilience triangle model (Bruneau et al., 2003). Achieving such performance-based resilience assessment requires a comprehensive evaluation of disruption impacts (Bi et al., 2025), thereby advancing the understanding of road system resilience under heavy rainfall events.

A meaningful assessment of road system resilience to heavy rainfall, or any other hazard, requires quantifiable metrics that align with the performance requirements and services central to the mission of transport agencies (National Academies of Sciences, Engineering, and Medicine, 2021). Yet, well-established performance metrics specifically designed to capture the multidimensional impacts of rainfall-induced acute road disruptions remain limited in transport operations (National Infrastructure Commission, 2024). A review of relevant reports published by the UK transport industry and government agencies (Department for Transport, 2020; Greater Manchester Transport Committee, 2023; National Highways, 2024, 2023; Office of Rail and Road, 2024a, 2024b; Transport for Greater Manchester, 2024; Transport for London, 2024; Transport Scotland, 2016) indicates that existing metrics primarily focus on capturing physical asset condition (e.g., percentage of pedestrian and cycle facilities in poor or very poor condition), operational performance (e.g., annual delay or journey time reliability), safety impact (e.g., the number of people killed or seriously injured recorded each year), environmental impact (e.g., noise, biodiversity, air quality, or carbon emissions), and user satisfaction (e.g., percentage of road users very or fairly satisfied in the national survey). Taken together, most existing metrics in transport operations are designed to monitor or evaluate road network performance under typical annual weather conditions, rather than to stress-test systems against extreme weather conditions and quantify the impacts of corresponding disruptions.

Following this, a literature investigation was conducted in the Web of Science database, using a search query combining keywords related to road transport, performance metrics, and rainfall. The result shows that existing studies tend to abstract road transport systems into networks and evaluate the effects of disruptions merely through changes in structural connectivity (Li et al., 2026; Liu et al., 2026a; Wu et al., 2026), using topological attributes such as average shortest path length (Bhatta et al., 2024; He et al., 2024), node degree (Li et al., 2026; Testa et al., 2015), global efficiency (Du et al., 2024; Haritha and Anjaneyulu, 2024), giant connected component (Wang et al., 2026; Wang et al., 2019), as well as centrality measures (Bhatta et al., 2024; Casali and Heinimann, 2019; Gangwal et al., 2023; Lu et al., 2024). Such topology-based analyses fail to capture performance and service-related aspects of road systems, providing information of limited practical relevance for understanding real-world disruption impacts. While increased travel time and speed change are adopted as performance metrics in literature (Arabi et al., 2023; Azargoshasbi et al., 2026; Dong et al., 2023; He et al., 2026; Li et al., 2026; Liu et al., 2025; Mathew and Pulugurtha, 2022; Pregnolato et al., 2017; Pyatkova et al., 2019; Zhang et al., 2026), the broader socio-economic consequences of disruptions are often overlooked in current assessment frameworks, despite the central role of road transport in supporting essential social and economic activities. As such, the full cost of any disruption to society is likely to be underestimated in analysis.

To evaluate these performance metrics for road transport resilience assessment, existing studies have developed a range of quantitative modelling methods. For instance, network model-based simulations are widely adopted to represent transport systems as node–link networks and assess resilience through changes in the above-mentioned topological attributes, operational performance indicators, or resilience properties such as redundancy (Xu et al., 2021) under disruption scenarios (Bi et al., 2023). While network-model based approaches are generally computationally efficient and suitable for large-scale system analysis, they are often limited in their ability to represent how individual travellers' behaviours and interactions collectively reshape traffic conditions, such as creating localised congestion under disruption (Liu et al., 2026a).

Traditional transport models, such as the four-step transport model, can simulate localised traffic congestions but are limited in their ability to capture flood-induced disruptions. Firstly, the spatio-temporal interactions among transport users, both within the transport network and across transport modes, are rarely considered (Azargoshasbi et al., 2026; Liu et al., 2026b; Ma et al., 2025; Mukesh et al., 2022; Tang et al., 2024; Wang et al., 2025; Xu et al., 2022; Zhang et al., 2023). Secondly, human behaviours are usually analysed in a highly aggregated manner, taking into account only specific and standard population groups (e.g., car users and commuters), rather than capturing a broader and more heterogeneous representation of the population with different socio-demographic attributes (e.g., age, gender, income, and occupation) and activity plans (e.g., trip purposes, departure times, and activity locations) (Álvarez Castro, 2024; Johnston, 2004; Li et al., 2025; Mladenovic and Trifunovic, 2014). Thirdly, key characteristics of the built environment, such as road types, number of lanes, maximum speed, permitted transport modes, and road capacity, are scarcely considered (Allalou et al., 2026; Franco et al., 2020; Manley et al., 2021). These three aspects are fundamental, as flooding can (1) limit interactions of transport users on the network as a result of the level of standing rainfall water in different places along the roads; (2) impact population groups differently depending on their socio-demographic attributes, where and when they travel, or the different modes of transport they use (e.g. private, public, or active modes); and (3) reduce permitted maximum speeds and flow capacity per road link, due to safety concerns and physical blockages by floodwater.

These limitations highlight the need for a modelling approach that can capture individual-level traveller behaviour, heterogeneous population characteristics, and dynamic network interactions under flood-induced disruptions. Agent-based models (AgBMs) offer such capability by representing individual travellers as autonomous agents with distinct activities, route choices, and adaptive behaviours (Railsback and Grimm, 2019; Shang et al., 2020; Stroeve and Everdij, 2017). It enables the simulation of dynamic interactions between travellers, infrastructure, and operational conditions under varying scenarios (Kim et al., 2022; Sun et al., 2021). This capability is especially relevant for heavy rainfall events, where disruption impacts are shaped not only by physical flood conditions but also by how travellers respond to changing network conditions and which groups are affected (Liu et al., 2025b). Accordingly, agent-based modelling has strong potential to capture system-wide resilience, offering opportunities to bridge physical disruption processes with broader operational, social, and economic consequences.

To address the limitations in performance metric design and advance road flood resilience assessment using AgBMs, this study (a) proposes multidimensional performance metrics to quantify flood disruption impacts and capture system resilience, covering physical, operational, environmental, social, and economic (POESE) perspectives; and (b) develops an integrated modelling framework that couples heavy rainfall modelling, spatially distributed surface water flood modelling, and agent-based multimodal road transport modelling to simulate rainfall-induced disruptions and quantify resilience with selected POESE metrics. Three behavioural scenarios are simulated, including a worst-case scenario where individuals are unable to adapt their behaviours, an ideal response scenario where individuals iteratively adjust their activity plans to avoid flooded road links, and an early warning scenario where a subset of individuals cancels trips in advance in response to flood warnings. The methodology is demonstrated through a case study of the Tyne and Wear region in North East England under a 1-in-30-year heavy rainfall event. This study contributes to the field of transport resilience and adaptation in the following ways:

(a) Positioned at the hazard-transport-society nexus, the proposed POESE performance metrics advances a systemic perspective in transport resilience assessment by tracing ripple effects of physical asset-level disruptions caused by flooding through operational performance loss towards broader environmental, social, and economic consequences. It offers a structured and comprehensive reference for quantifying transport resilience to heavy rainfall across multiple dimensions, enabling clearer interpretation of disruption impacts and more consistent benchmarking across events.
(b) The integrated modelling framework represents an advanced, interoperable synthesis of rainfall, flood, and agent-based transport simulations, enabling spatially dynamic analysis of disruption propagation, travel behavioural responses, and systemic performance under adverse weather conditions. This integrated approach provides a structured protocol for conducting future evidence-based assessment of transport resilience, offering a means for stress-testing transport systems under a range of extreme weather scenarios and evaluate the effectiveness of adaptation strategies.

The reminder of the paper is organised as follows. Section 2 presents the proposed POESE performance metrics for road flood resilience assessment. Section 3 describes the agent-based integrated modelling framework designed to support the assessment of selected POESE metrics, along with introducing the case study. The assessment results are presented in Section 4, followed by a discussion in Section 5. Conclusions are summarised in Section 6.

## 2. POESE performance metrics for flood resilience assessment

The POESE performance metrics are captured from analysing road disruption impact chains (Wang et al., 2025) following heavy rainfall events. They trace how disruptions at physical road asset level propagate through operational performance to broader environmental, social, and economic consequences.

### *2.1 Flood disruption impact chains and performance metrics*

Fig. 1 presents the impact chains of long-term road disruptions caused by heavy rainfall, covering both immediate disruptions during the event and the prolonged recovery process due to catastrophic asset damage. The disruption begins with flooding on road links, leading to reduced maximum vehicle speeds of road links or, in cases of deep water or severe asset damage, road closures that trigger vehicle rerouting. This physical asset-level disruption is reflected in reduced road network availability (Panakkal and Padgett, 2024), which can be measured by the number and total length of road links affected by flooding with varying depths (represented by box PH1-3 in Fig. 1; metrics PH1-3 are detailed in Table 1). Either speed reduction or rerouting results in journey delay, which is a measure of operational performance loss. Its significance can be evaluated using total delay, average delay, or peak-time average delay (see metrics OP1-4). Subsequently, traffic congestion and longer travel distances from rerouting increase greenhouse gas (GHG) and air pollutant (e.g., $NO_x$ and $PM_{2.5}$) emissions (Chen et al., 2022; Wang et al., 2025), which serve as a measure of the environmental impact of disruption (metrics EN1-3). Following this, the rise in air pollutants during long-term disruptions could be substantial and lead to a greater burden of disease such as lung cancer, asthma, and stroke (Litman, 2013; Woodcock et al., 2009). This metric reflects the induced health issue of disruptions (metric SO7), which is a measure of social impact.

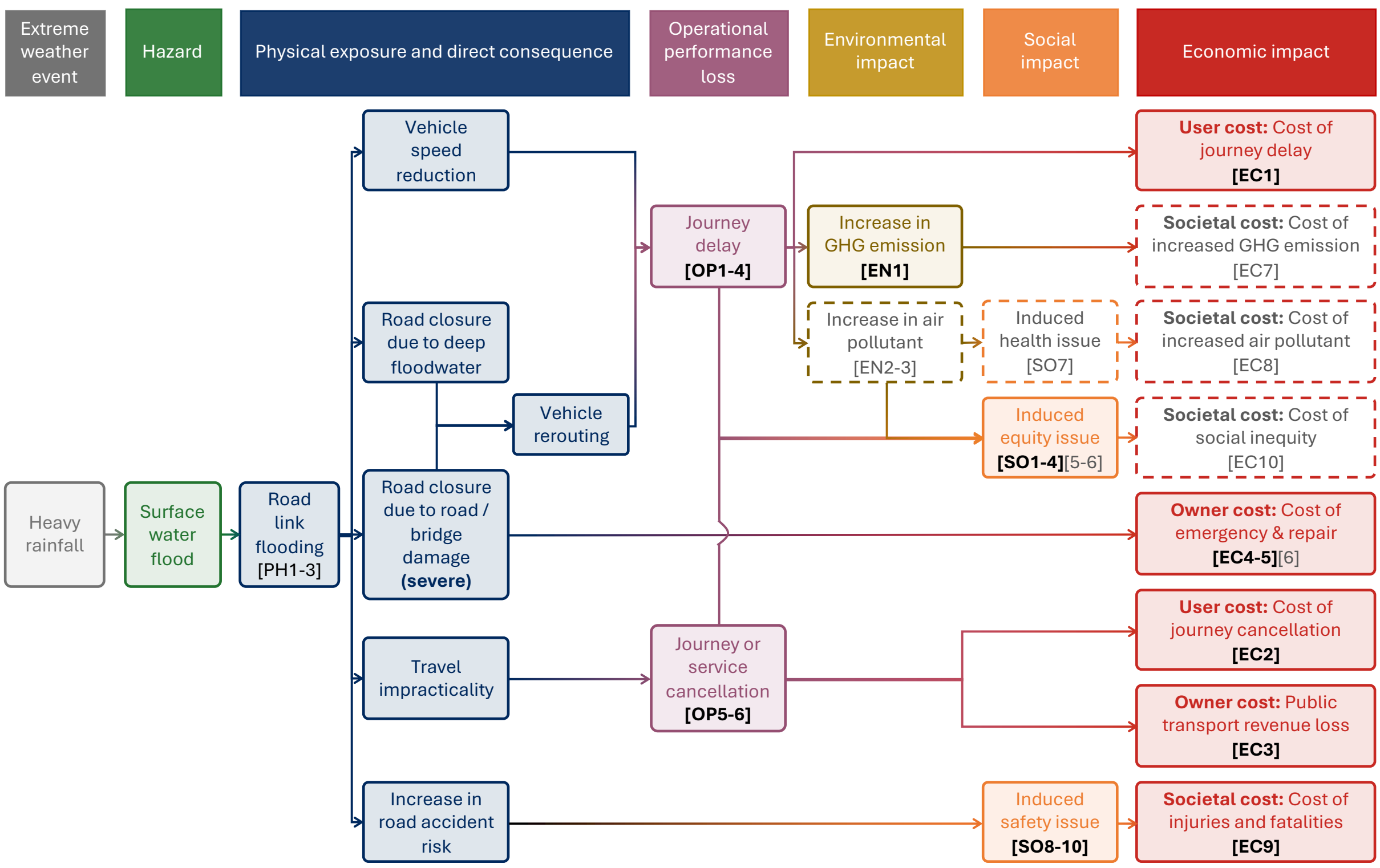

**Fig. 1** Impact chains of road disruptions with catastrophic asset damage and prolonged recovery following heavy rainfall (explanations of the codes in black and grey are provided in Table 1)

When feasible travel routes become unavailable, transport users are forced to cancel their journeys, and bus services may also be suspended (metrics OP5-6 in Table 1). Whether in the form of journey delay or cancellation, operational performance loss affects users' access to employment, education, healthcare, and other essential services. These impacts could be unevenly distributed, with vulnerable groups experiencing a greater burden (Coleman et al., 2024; Soriano et al., 2022). As such, assessing the disproportionate delay and cancellation effects across population groups and identifying their geographical distributions (metrics SO1-4) is essential for incorporating equity into resilience assessment (Coleman et al., 2024), enabling adaptation strategies to explicitly target vulnerable populations. This decreased mobility represents another measure of social impact. Analysing decreases in accessibility of different population groups (metric SO5) is particularly relevant for long-term disruption (Miller, 2022; Mitropoulos et al., 2025), as sustained limitations could exacerbate social inequity. Another inequality aspect is that the increased air pollutant emissions may concentrate in specific spatial areas, disproportionately affecting residents in those locations (metric SO6). Finally, road flooding increases traffic accident risk due to reductions in visibility and vehicle control (Theofilatos and Yannis, 2014), as well as traffic rerouting from safer major roads to less suitable minor residential streets, which may induce additional safety issue that is usually measured by the number of casualties and fatalities (metrics SO8-10) (Sattar et al., 2024).

**Table 1** POESE performance metrics and their applicability

| Code | Performance metrics | A. Short-term disruption | B. Long-term disruption |
|---|---|---|---|
| PH1* | Number of road links affected by flooding | ✓ | ✓ |
| PH2* | Total length of road links affected by flooding (km) | ✓ | ✓ |
| PH3 | Total length of critical roads (e.g., A-Roads in the UK) affected by flooding at varying depths (km) | ✓ | ✓ |
| OP1* | Total delay (min) | ✓ | ✓ |
| OP2* | Average delay (min) | ✓ | ✓ |
| OP3* | Morning-peak (7-9 AM) average delay (min) | ✓ | ✓ |
| OP4* | Afternoon-peak (4-6 PM) average delay (min) | ✓ | ✓ |
| OP5* | Number of cancelled journeys by trip purpose | ✓ | ✓ |
| OP6 | Number or percentage of cancelled bus service | ✓ | ✓ |
| EN1* | Increase in GHG emission (kgCO2e) | ✓ | ✓ |
| EN2* | Increase in $NO_x$ emission (kg or $\mu g/m^3$) | | ✓ |
| EN3* | Increase in $PM_{2.5}$ emission (kg or $\mu g/m^3$) | | ✓ |
| SO1* | Percentage of individuals experiencing delayed journeys by demographic group (%) | ✓ | ✓ |
| SO2* | Percentage of individuals experiencing cancelled journeys by demographic group (%) | ✓ | ✓ |
| SO3 | Geographical distribution of individuals experiencing delayed journeys by demographic group | ✓ | ✓ |
| SO4 | Geographical distribution of individuals experiencing cancelled journeys by demographic group | ✓ | ✓ |
| SO5 | Decrease in the number or percentage of a population group who are able to access a type of destination from home within a given proximity by a specified transport mode | | ✓ |
| SO6 | Geospatial distribution of increased air pollutant emission across demographic groups ($\mu g/m^3$) | | ✓ |
| SO7 | Increased burden of disease | | ✓ |
| SO8 | Number of slightly injured individuals | ✓ | ✓ |
| SO9 | Number of seriously injured individuals | ✓ | ✓ |
| SO10 | Number of fatalities | ✓ | ✓ |
| EC1* | Cost of journey delay (£) | ✓ | ✓ |
| EC2* | Cost of journey cancellation (£) | ✓ | ✓ |
| EC3 | Public transport revenue loss (£) | ✓ | ✓ |
| EC4 | Cost of emergency response (£) | ✓ | ✓ |
| EC5 | Cost of asset repair (including labour costs) (£) | ✓ | ✓ |
| EC6 | Cost of asset reconstruction (including labour costs) (£) | | ✓ |
| EC7* | Cost of increased GHG emission (£) | | ✓ |
| EC8* | Cost of increased air pollutant emission (£) | | ✓ |
| EC9 | Cost of injuries and fatalities (£) | ✓ | ✓ |

| EC10 | Cost of social inequity (£) | ✓ |
|---|---|---|

Note: * denotes metrics that are selected for demonstration in the case study. Metrics without an asterisk are not assessed in this case study, as their evaluation would require additional model components, data inputs, and dedicated analysis beyond the intended scope of this framework demonstration.

The above consequences can be monetised to reflect the economic impact of disruptions (Contreras-Jara et al., 2025), providing a potential synthesis index of system resilience that has practical meaning, allows for comparison across events, and forms the basis of cost-benefit analysis of potential adaptation interventions. As indicated in Fig. 1, economic impact consists of user costs, owner costs, and societal costs. User costs refer to the direct burden experienced by individuals, including the cost of journey delay and cancellation. The cost of journey delay (metric EC1 in Table 1) can be quantified as the product of total delay time and the value of time, which should ideally be differentiated by journey purpose to account for variation in the economic value of time across different travel purposes (e.g., commuting, business, or leisure). The cost of journey cancellation (metric EC2) can be estimated based on the number of cancelled trips and the associated generalised travel cost, reflecting the welfare loss incurred when trips cannot be completed. To the best of the authors' knowledge, empirical estimates of cancellation costs are currently unavailable. Therefore, the generalised travel cost may be approximated using proxy measures such as income-based estimates or by applying a fixed number of hours valued using the value of time.

Owner costs encompass the financial losses of entities that own or manage transport systems, including revenue loss from bus service cancellations (metric EC3, which can be obtained through ticket loss (Bi et al., 2024)), as well as the costs of emergency response (metric EC4), asset repair (metric EC5), and asset reconstruction (metric EC6) resulting from road asset damage. These costs can be informed by collecting historical event data. Finally, societal costs capture the broader external impacts on the wider community, including those arising from increased emissions (metrics EC7-8), injuries and fatalities (metric EC9), and social inequity (metric EC10). Unit costs for emissions, injuries, and fatalities can be obtained from established transport appraisal guidance (Department for Transport, 2024). The cost of social inequity, however, is inherently difficult to quantify, reflecting the absence of widely accepted methods for valuing distributional and equity-related impacts. Where possible, economic metrics with available unit cost data should be adopted first as a starting point. As methodological approaches mature and supporting data become more robust, the ability to quantify a broader set of the proposed metrics is expected to improve, enabling more comprehensive economic impact assessments.

### *2.2 Applicability of metrics in short-term disruptions*

The above impact chains and performance metrics are analysed under long-term disruption scenarios characterised by catastrophic asset damage and extended recovery periods. In the case of short-term disruptions without prolonged recovery, such as disruptions triggered by rainfall but fully restored within 24 hours or several days, some impacts are expected to be minimal and can therefore be excluded from the assessment. As presented in Table 1, the increase in air pollutant emissions (metrics EN2-3) within a short period is likely to be negligible. Therefore, analysing the geospatial distribution of increased air pollutant emissions across demographic groups (metric SO6) is less meaningful for individual short-disruption events. Additionally, the costs associated with increased GHG and air pollutant emissions (metrics EC7-8) are likely to be minimal and may not warrant explicit quantification. Furthermore, evaluating the decrease in general accessibility (metric SO5) for short-term disruptions is usually of limited significance, as such brief interruptions do not alter the overall accessibility pattern. Nevertheless, accessibility losses remain highly relevant for healthcare and emergency services, where even short-lived disruptions can have undesirable consequences (Yang et al., 2022). Finally, the cost of asset reconstruction (metric EC6) is not considered, as no reconstruction is typically required for short-term disruptions.

It should be noted that the long-term and short-term indicator sets are not intended to be combined through weighting or aggregated into a single composite resilience score. Rather, they provide context-specific assessment options, from which relevant metrics can be selected according to the modelling purpose, damage severity, recovery duration, and data availability.

**3. Agent-based integrated modelling framework with case descriptions**

To demonstrate the applicability of the proposed metrics, an agent-based integrated modelling framework is developed to simulate flood-induced road disruptions following heavy rainfall and quantify multidimensional impacts. As presented in Fig. 2, the framework begins with the simulation of rainfall events to generate scenarios that capture the intensity, duration, and spatiotemporal distribution of heavy rainfall. These outputs feed into an urban flood model to produce high-resolution, spatially distributed surface water flood depth maps at regular time intervals (e.g., 20 minutes), indicating the temporal evolution of flood conditions from event onset to recession. The resulting flood depth maps then inform vehicle travel speed constraints in an agent-based transport model to stress-test the road network, identifying road segments affected by inundation and simulating the adaptive behaviour of individual agents. This simulation captures emergent individual travel disruptions caused by flooding, such as journey delay and cancellation, which are then evaluated through a set of selected performance metrics proposed above. The remainder of this section details the process of each module of this framework, illustrated through the case study of the Tyne and Wear region in North East England (Fig. 3). This region is selected because it is the main urban conurbation in the North East, with approximately one million inhabitants, representing around 40% of the region's population.

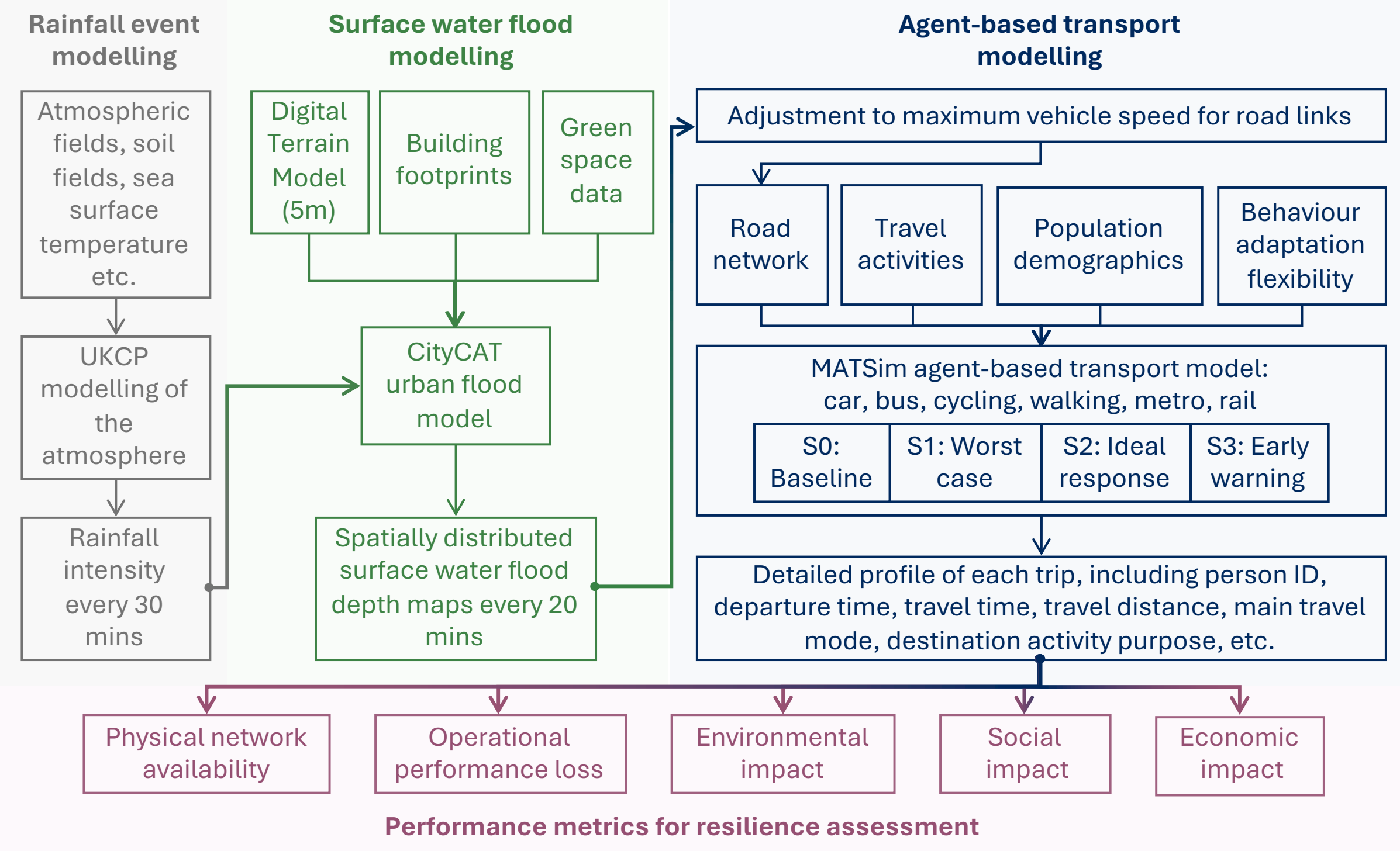

**Fig. 2** Integrated modelling framework with performance metrics

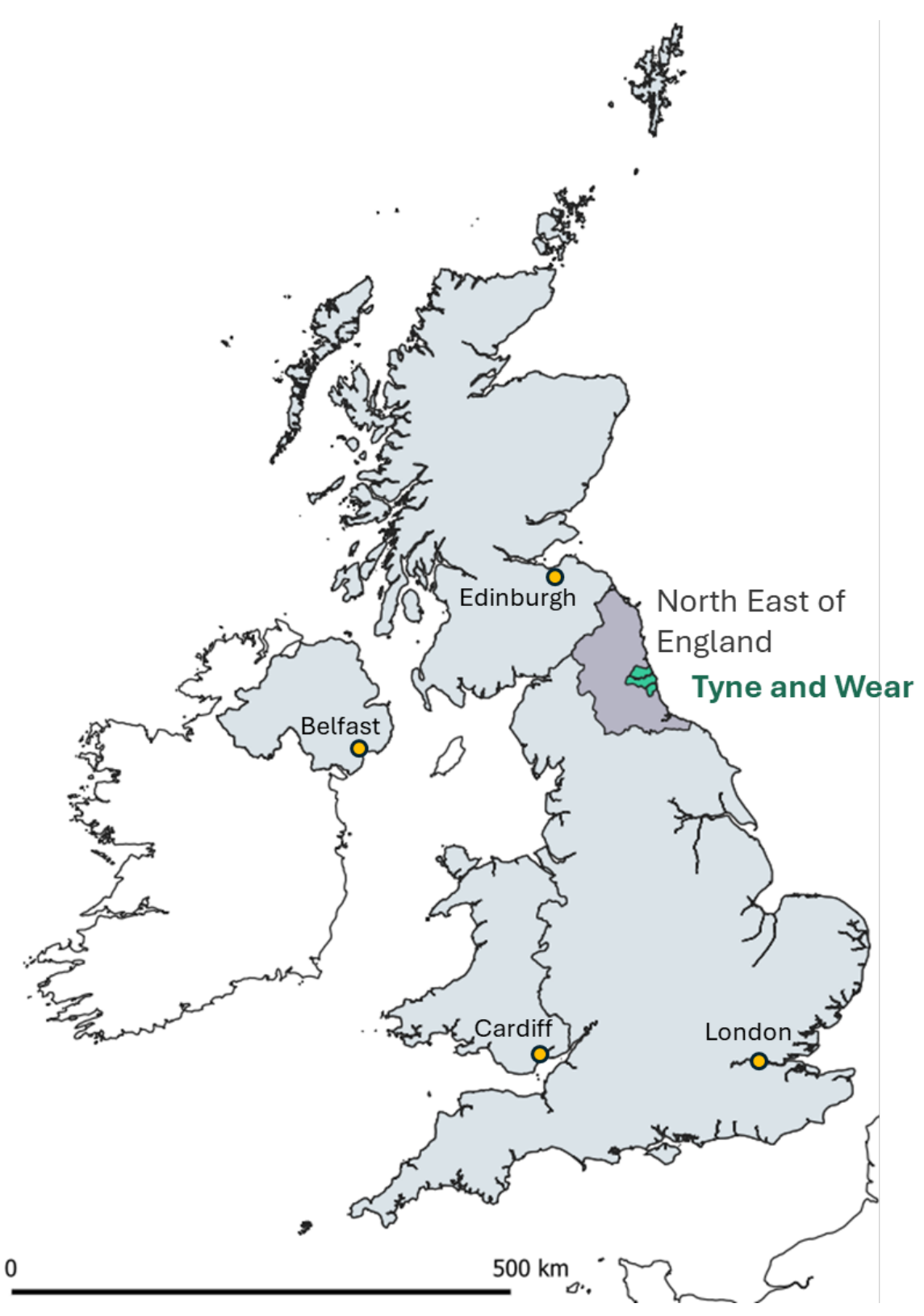

**Fig. 3** Location of Tyne and Wear in North East England

### *3.1 Rainfall event modelling*

To provide modelled input data for rainfall events, this study utilises heavy rainfall events extracted from the UK Climate Projections (UKCP18) (Met Office, 2018). Specifically, the event is derived from the standard member of the UKCP18-Local ensemble, corresponding to the current climate. The UKCP18-Local comprises twelve 100-year-long transient simulations at a ~2.2-km horizontal resolution, conducted at the UK Met Office Hadley Centre (Kendon et al., 2021; Kendon et al., 2020). This rainfall model output has mean precipitation rate in 30-min intervals, which are further processed to cut-out Tyne and Wear and calculate corresponding statistics for the region, as illustrated in Fig. 4. The selected event represents a 1-in-30-year heavy convective rainfall (defined in terms of sub-hourly rainfall maxima) in the native grid, with a maximum of 109.8 mm/h precipitation intensity in the study area. This is a highly localised short-duration heavy rainfall event with flash flood potential, affecting the region for three hours.

### *3.2 Surface water flood modelling*

Based on the rainfall modelling outputs, CityCAT is adopted to produce dynamically and spatially distributed surface water flood depth maps. CityCAT is an innovative hydrodynamic urban flood modelling system developed by Newcastle University (Glenis et al., 2018). The software enables rapid assessment of combined pluvial and fluvial flood risks and effectively considers the influence of ground elevations, buildings, green spaces, and flood intervention measures. Leveraging advanced object-oriented programming architecture and high-resolution numerical methods, CityCAT can efficiently and accurately handle complex urban flood scenarios, including those across large metropolitan areas.

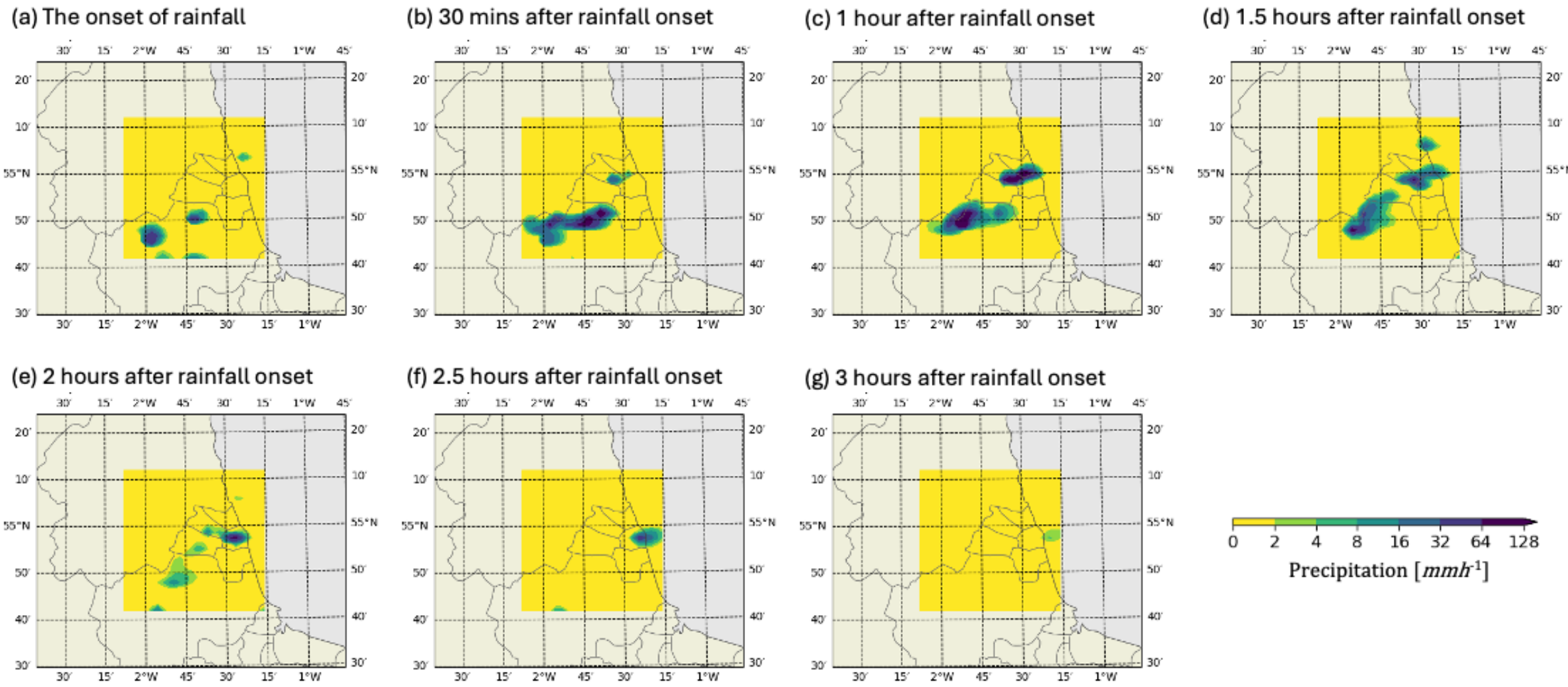

**Fig. 4** Spatial distribution of rainfall intensity at different time points following rainfall onset

In terms of data demand, CityCAT primarily requires underlying land surface data and rainfall inputs. Rainfall inputs can be either synthetic design storms or observed precipitation events, supporting applications in both planning and emergency response. Underlying surface information is typically derived from a Digital Terrain Model (DTM), along with topographic spatial data on building footprints and green spaces—used to classify impervious and pervious surfaces. This allows the model to represent urban hydrology with spatial precision, accounting for surface runoff pathways, barriers to flow, and infiltration capacity. To account for the presence of buildings, the model grid adopts the Building Hole approach, treating building areas as impermeable zones by encircling them with non-flow boundaries, effectively removing them from the active flow domain. CityCAT has been validated through three representative cases, including an analytical solution in a parabolic bowl, a dam-break physical experiment, and a real-world urban flood scenario (Glenis et al., 2018). This demonstrates the model's capability in assessing urban flood risk and supporting city-scale decision-making.

To understand how heavy rainfall generates surface water flooding that subsequently affects road transport, a CityCAT flood model is set up for Tyne and Wear using a 5-m DTM, building footprints, and green space data, which are sourced from Ordnance Survey open datasets (Digimap, 2020). The roughness coefficient (Manning's) is set to 0.02 for the impermeable surfaces and 0.035 for the permeable surfaces (Iliadis et al., 2023). Outputs from the 1-in-30-year heavy rainfall modelling in the previous step are processed as a spatially differentiated input across the entire domain over a 3-hour duration. The flood simulation is run for a total duration of six hours, including a 3-hour post-rainfall period. Surface water flood depth maps are produced at 20-min intervals, which capture the spatio-temporal evolution of flooding across the study area and serve as input data for the subsequent transport model. Fig. 5 illustrates the progression of flooding, with water depths mapped on a 5-m grid.

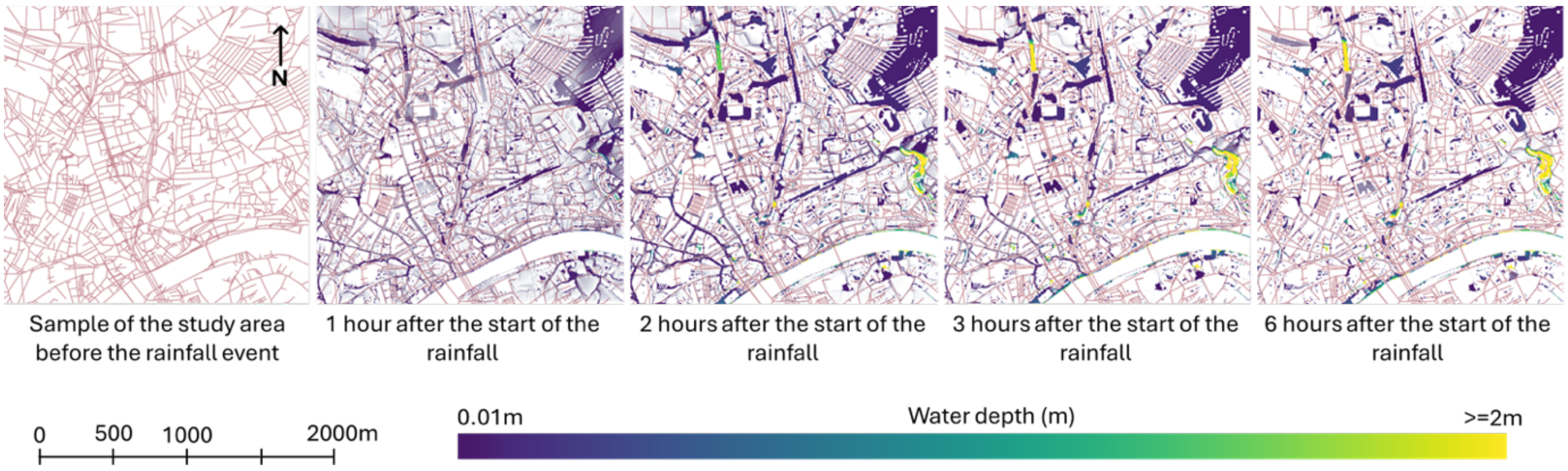

**Fig. 5** Spatio-temporal evolution of surface water flooding across the study area

### *3.3 Agent-based transport modelling*

#### *3.3.1 Baseline scenario (S0)*

To simulate how surface water flooding disrupts road passenger transport, agent-based modelling is applied. In the case of transport AgBMs, agents can be represented by individuals, and the environment by the transport network and public transport services. Each agent is characterised by a set of socio-demographic attributes and a daily travel activity plan, which together define who they are, what they do, and where they do it. The transport network is defined by the characteristics of individual road links, which enable agents to move and interact in space and time according to their activity plans, alongside the different public transport services in the region (i.e., routes, stops, and service types). These models provide a bottom-up approach, where the spatiotemporal interactions between agents and the environment are considered first, intending to investigate the emergent and collective effects on the system (Huang et al., 2022; Moyo Oliveros and Nagel, 2016).

As shown in Fig. 6, the Tyne and Wear case study is developed based on the multi-modal transport MATSim (Horni et al., 2016) AgBM, which was validated in the authors' previous work (Alvarez Castro et al., 2024). This model covers cars, buses, metros, rail, cycling, and walking, and simulates travel activities of 20% the regional population (approximately 200,000 agents) over the 24 hours of a normal day. Population subsampling is a standard approach in large-scale agent-based transport modelling to manage computational complexity and execution time. In this study, the 20% sample is higher than that typically adopted in existing literature and doubles the minimum recommended ratio (Mersini and Ziemke, 2026), thereby improving population representation while maintaining computational feasibility. Accordingly, the characteristics of the road network (e.g., flow capacity, flow storage) were also adjusted to 20% to maintain consistency between the components of transport demand and supply.

Each agent is characterised by 10 socio-demographic attributes (i.e., age, sex, marital status, child dependency, economic activity, occupation, annual gross income, driving license, car access, and bicycle access) and a detailed activity plan (i.e., trip purpose, departure time, transport mode used, and origin and destination locations), using datasets from UK Census (Office for National Statistics, 2026a), Office for National Statistics (Office for National Statistics, 2026b) and National Travel Survey (NTS) (Department for Transport, 2026). Activity plans were assigned to agents based on common socio-demographic attributes with NTS surveyed individuals, providing different mobility patterns (e.g., number of trips, destinations, trip purposes, departure times, and travel modes) based on common characteristics. This approach assumes that people with similar characteristics behave similarly during a normal working day. During the simulation, all agents follow the same rules, with the same three types of strategies throughout the MATSim co-evolutionary algorithm (Horni et al., 2016): (1) the possibilities of using different modes of transport (except cars, which are only allowed for those agents with a driver's license and access to a car), (2) changing the routes followed, and (3) adapting departure times. More details of the three different strategies are provided in Section 3.3.2. Trip destinations are fixed in all cases.

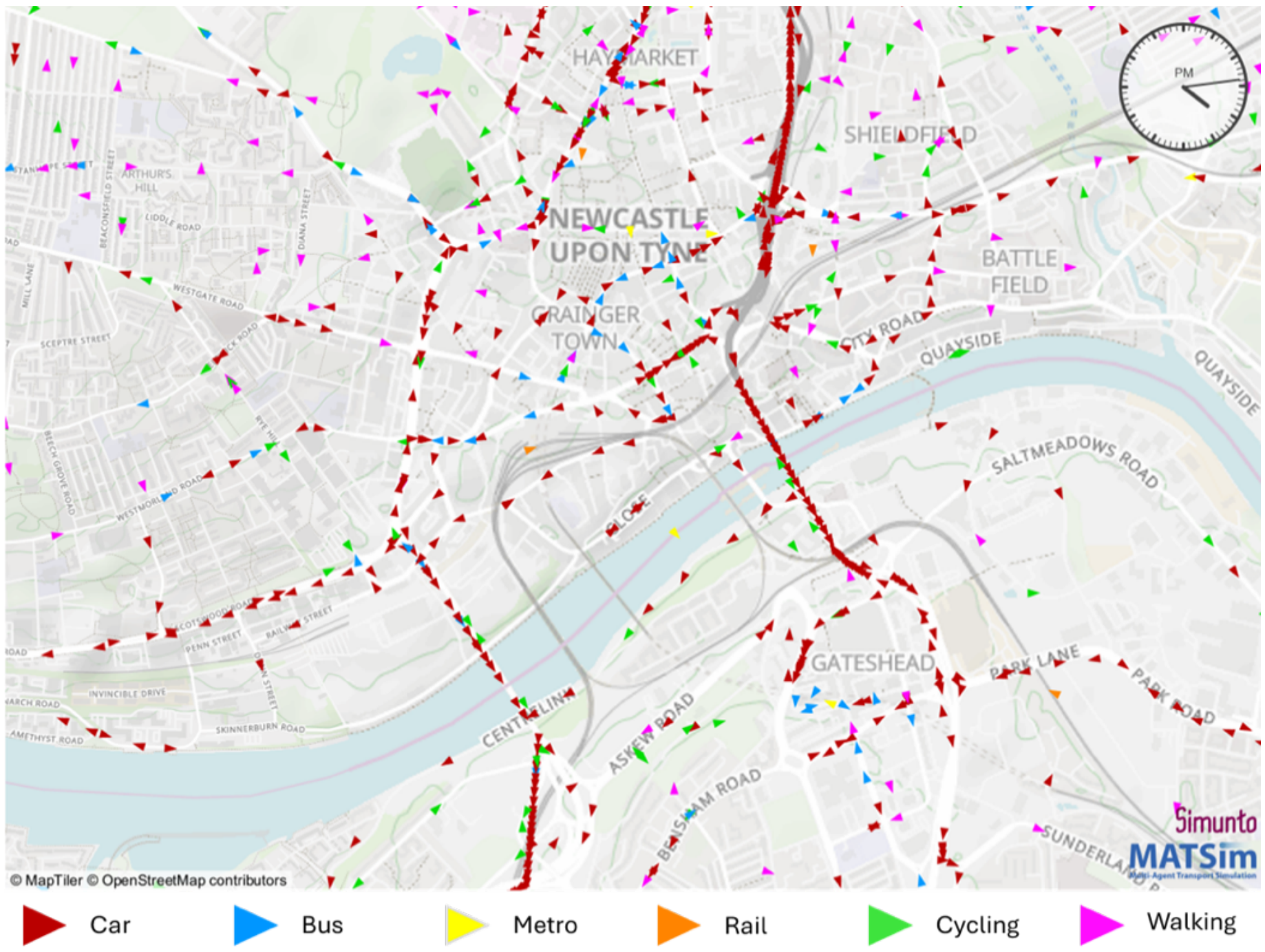

**Fig. 6** Sample representation of the multi-modal transport demand model in Tyne and Wear

The transport network is defined using OpenStreetMap's (OpenStreetMap, 2018) roads and public transport information (Rail Delivery Group, 2023; Traveline, 2023) from the study area, including characteristics such as permitted modes, maximum speed, maximum capacity, and details of public transport services. This simulated normal-day scenario serves as the baseline scenario (S0), against which the impacts of flood disruptions in other scenarios are evaluated. For a detailed description of the model development, calibration and validation, refer to (Alvarez Castro et al., 2024).

*3.3.2 Behavioural scenarios under flood disruptions (S1-S3)*

Building upon this baseline scenario, the set of surface water flood depth maps generated in the previous flood modelling step are utilised to adjust the maximum vehicle speed for each road link, thereby simulating the immediate asset-level impacts of flooding on the road network.

Time-dependant flood depth maps are intersected with the OpenStreetMap road network of the study area, using GIS software (e.g., QGIS). The maximum depths of standing water of each road link over time are extracted, which are then converted into maximum vehicle velocities based on the relationship between water depth and vehicle speed established by Pregnolato et al. (2017). This relationship is represented as a curve linking the depth of water on a road at a specific timeframe with the maximum allowable vehicle speed, under the assumption of safe driving behaviour. With reference to this curve, the model uses a threshold of 0.01 m to detect floodwater on road links, and 0.3 m to identify severely flooded road links where vehicle speed is adjusted to zero. These adjusted maximum vehicle speed limits are dynamically incorporated into the MATSim transport network (using a NetworkChangeEvents.xml file within the MATSim network config file functionality (Rieser et al., 2016), as shown in Fig. 7) every 20 mins, which is aligned with the flood depth map intervals. This functionality allows specifying in the model when (i.e. specific time of day in HH:MM:SS format), where (specific road links affected), and how (maximum free speed allowed in m/s) the effects of the extreme rainfall event impact the transport road links. Therefore, agents using those affected road links are disadvantaged, as their journeys take longer (i.e., lower vehicle speeds are permitted) or may not be able to be completed if the road links are considered completely flooded (i.e., with a water depth equal to or greater than 0.3 m).

```
<networkChangeEvent startTime="11:00:00">
        <link refId="1534890"/>
        <link refId="7530969"/>
        <link refId="398643"/>
        <freespeed type="absolute" value="2.4"/>
</networkChangeEvent>
<networkChangeEvent startTime="11:20:00">
        <link refId="1534890"/>
        <link refId="4657452"/>
        <link refId="9635146"/>
        <freespeed type="absolute" value="1.8"/>
</networkChangeEvent>
```

**Fig. 7** Short representation of the networkChangeEvents.xml file, including the road links (i.e., "link refId") impacted by the rainfall event with a maximum free speed value (i.e., the "freespeed" parameter in meters per second) at a specific time of the day (i.e., the "startTime" parameter in the HH:MM:SS format).

Due to the lack of literature on the impact of extreme rainfall on walking and cycling, the same speed restrictions are applied to all road modes. The rainfall event is set to begin at 7 AM and lasts for 3 hours, with an additional 3 hours included to capture the spatial and temporal evolution of the resulting flood. This allows the model to simulate the potential impact of flooding on road travel during the morning peak and throughout the rest of the day.

As summarised in Table 2, three agent behaviour scenarios are simulated in this study. The first scenario represents the worst case (S1), where agents are unable to adapt their behaviours and change routes. If an agent needs travel on roads affected by flooding, their journey is delayed because of the reduced maximum speeds on those road links. Furthermore, if a road link is severely flooded with water depth of 0.3 m or more, the agent can no longer continue their journey and remains at that location for the rest of the day, cancelling any subsequent trips. Scenario S1 is run for a single iteration.

The second scenario represents ideal response (S2), where agents learn and adapt their behaviours over a number of iterations, following the MATSim co-evolutionary algorithm (Horni et al., 2016). In each iteration, a random 30% of agents are allowed to modify their activity plans—such as switching to different transport modes, taking detours, or changing trip departure times—to identify, when possible, the most effective ways to avoid severely flooded or congested roads. After each iteration, daily agents' satisfaction (i.e., a utility, typically referred to as 'score' in MATSim) is measured applying the Charypar-Nagel scoring function (Charypar et al., 2016; Charypar and Nagel, 2005). Adapted activity plans with low satisfaction scores are discarded, while only the best-performing plans are retained (five per agent in this scenario). Scenario S2 is run for 500 iterations. It allows agents to explore, on average, 120 alterative activity plans during the first 80% of iterations. During the final 20% iterations, agents select from among their previously learned best activity plans to minimise the impact of extreme rainfall on their daily routines.

**Table 2** Characteristics of the four simulated scenarios

| Characteristic | S0: Baseline | S1: Worst case | S2: Ideal response | S3: Early warning |
|---|---|---|---|---|
| Weather condition | Normal | 30-year rainfall | Same as S1 | Same as S1 |
| Simulated population | 20% of the total population from the study area (208,512 agents) | Same as S0 | Same as S0 | 8% of the total population (82,588 agents) after assuming the following agents in S0 would stay at home:<br>• 100% of "students"<br>• 100% of "sick" agents with no medical appointments<br>• 0% of agents working in healthcare facilities<br>• 50% of "employed" agents<br>• 50% of "retired" agents<br>• 50% of "looking after home or family" agents |
| Travel demand | 592,081 trips simulated | Same as S0 | Same as S0 | 238,338 trips simulated |
| Iteration | / | 1 time | 500 times | Same as S2 |
| Allowable travel behaviour | / | None | 30% of random agents in every iteration are allowed to try the following alternative options:<br>• 10% reroute<br>• 10 % switch transport mode<br>• 10% change travel start time (±10 mins) | Same as S2 |

Lastly, the third scenario reflects a proactive response in the presence of an early weather warning system provided by the local authority (S3). In this case, a subset of agents is instructed to stay at home, thereby reducing the overall exposure to heavy rainfall-induced road disruptions. This subset of agents is selected based on their socio-demographic attributes, under the assumption that certain population groups (e.g., students or sick individuals without scheduled appointments) would cancel their trips in response to early flood warnings, while others (e.g., agents working in healthcare facilities) would continue with their planned journeys due to the essential nature of their roles or needs. For agents lacking detailed employment information, a neutral assumption is applied—50% are randomly assigned to stay at home—to reflect uncertainty in individual responses. These assumptions capture the different levels of flexibility and obligation across the population in adapting to flood-related disruptions. Similar to scenario S2, scenario S3 is also run for 500 iterations, with other settings remaining the same. The only difference is that it simulates 40% of the total population in all other scenarios, due to the early warning response.

The outputs from the MATSim AgBM for each scenario, as indicated in Fig. 8, provide a detailed description of the agents' updated activity plans, such as main transport mode, departure time, travel distance, trip purpose, and origin-destination locations, showing how agents adapt their behaviour, when possible, in response to an extreme rainfall event. Comparisons between the later scenarios and the baseline scenario illustrate how agents face the rainfall event for the first time without the ability to adapt (S1), how they manage to avoid affected roads when possible (S2); and how the early warning helps reduce the proportion of the population exposed to flood risk and disruptions (S3).

| Person ID | Departure time | Travel time | Waiting time | Travel distance | Main mode | Origin activity | Destination activity | Origin coordinate | Destination coordinate |
|---|---|---|---|---|---|---|---|---|---|
| Person_BB | 08:17:52 | 00:29:42 | 00:00:00 | 990 m | walk | home | education | 433925, 570002 | 434347, 570898 |
| Person_BB | 17:24:18 | 00:13:21 | 00:01:16 | 990 m | pt | education | home | 434347, 570898 | 433925, 570002 |
| Person_CC | 08:01:42 | 00:09:57 | 00:00:00 | 2985 m | car | home | work | 439182, 556451 | 436966, 554450 |
| Person_CC | 11:55:09 | 00:07:51 | 00:00:00 | 1761 m | car | work | supermarket | 436966, 554450 | 436029, 555942 |
| Person_CC | 12:49:53 | 00:09:08 | 00:00:00 | 3193 m | car | supermarket | home | 436029, 555942 | 439182, 556451 |
| Person_DD | 08:38:22 | 00:11:29 | 00:00:00 | 1573 m | bike | home | education | 433686, 569908 | 435259, 569913 |
| Person_DD | 16:42:17 | 00:10:13 | 00:00:00 | 1573 m | bike | education | home | 435259, 569913 | 433686, 569908 |
| Person_DD | 17:23:26 | 00:07:52 | 00:00:00 | 371 m | walk | home | leisure_act | 433686, 569908 | 433573, 569554 |
| Person_DD | 17:45:12 | 00:09:43 | 00:00:00 | 371 m | walk | leisure_act | home | 433573, 569554 | 433686, 569908 |

**Fig. 8** Example of outputs from the MATSim AgBM

### *3.4 Metrics selected for demonstration in the case study*

Based on the modelling capacity and the initial AgBM outputs outlined in Fig. 8, sixteen metrics are selected for assessing disruption impacts of the case study. As indicated in Table 1, the selected metrics consist of two physical, five operational, three environmental, two social, and four economic performance metrics. Specifically, reduced network availability (PH1-2) is derived from GIS software (e.g., QGIS) by overlaying the road network with flood maps. Travel details such as time and distance are obtained directly from AgBM outputs to support further analyses of delays (OP1-4), cancellations (OP5), emissions (EN1-3), equity (SO1-2), and economic costs (EC1-2 and EC7-8). Although the case study focuses on short-term disruptions, emissions are still measured to check the earlier assumption that they may be negligible for short-term disruptions. Increased emissions are calculated using a simple measure: the increased travel distance multiplied by the corresponding emission factor (HM Government, 2019) (available from (Department for Energy Security and Net Zero, 2025; National Atmospheric Emissions Inventory, 2023)). Induced equity issues are measured based on the proportion of affected agents in each population group, disaggregated from age, gender, employment status, and income. User and societal costs are then estimated by applying the relevant unit costs, provided by the UK Department for Transport (Department for Transport, 2024), to delay hours, the number of cancelled journeys, and increased emissions.

It is worth noting that the costs of journey delay and cancellation are differentiated by trip purpose, due to varying values of time associated with different travel activities. The values of time for different travel purposes are provided as £12.86/h assigned for commuting and £5.87/h for other purposes, with suggested sensitivity test ranges of ±25% and ±60%, respectively (Department for Transport, 2024). Informed by this, the value of time is set at £12.86/h for work trips, £9.39/h for medical trips (+60% based on £5.87/h), £7.63/h for education trips (+30% based on £5.87/h), and £5.87/h for all other trips.

In terms of the values of cancelled journeys by travel purpose, to the best of the authors' knowledge, these have not yet been properly studied. In the absence of any guidance, it is assumed that the cost of a cancelled journey is equivalent to eight hours' worth of the value of time associated with each respective trip purpose, including medical, work, education, and home. For shopping, leisure, and other trip purposes, journey cancellation costs are not considered due to their lower criticality during short-term disruptions.

The remaining unselected metrics, mainly social and indirect economic metrics, would require additional model components, data inputs, and dedicated methodology design and analysis. For instance, quantifying flood-induced transport injuries or fatalities (metrics SO8-10) would require coupling the transport disruption model with pedestrian exposure, evacuation, and casualty estimation models, supported by detailed information on passenger locations, evacuation behaviour, hazard intensity, and emergency response conditions. Similarly, estimating the cost of social inequity (metric EC10) would require dedicated methodological development, as no established approach is readily available for monetising inequity arising from transport flood disruption across population groups in this context. Therefore, such metrics are left for future modelling development rather than demonstrated in this case study.

## 4. Results

Building on the above methodology and dataset, this section presents assessment results of rainfall impacts for scenarios S1, S2, and S3 in comparison to the baseline scenario S0.

### *4.1 Physical network availability*

As depicted in Fig. 9, the number and total length of road links affected by flooding are measured over time. Within the first two hours of heavy rainfall, 48,844 road links experience at least 0.01 m of water depth, accounting for 15.3% of the total network. The peak of severely flooded roads with water depth above or equal to 0.3 m is observed during 9:20-9:40, with 2,645 road links closed, accounting for 0.8% of the total. In terms of total affected road length, links with at least 0.01 m floodwater depth reach the peak during 9:20-9:40, with 3,433 km affected (13.2% of the total), while 226 km (0.9%) are considered closed.

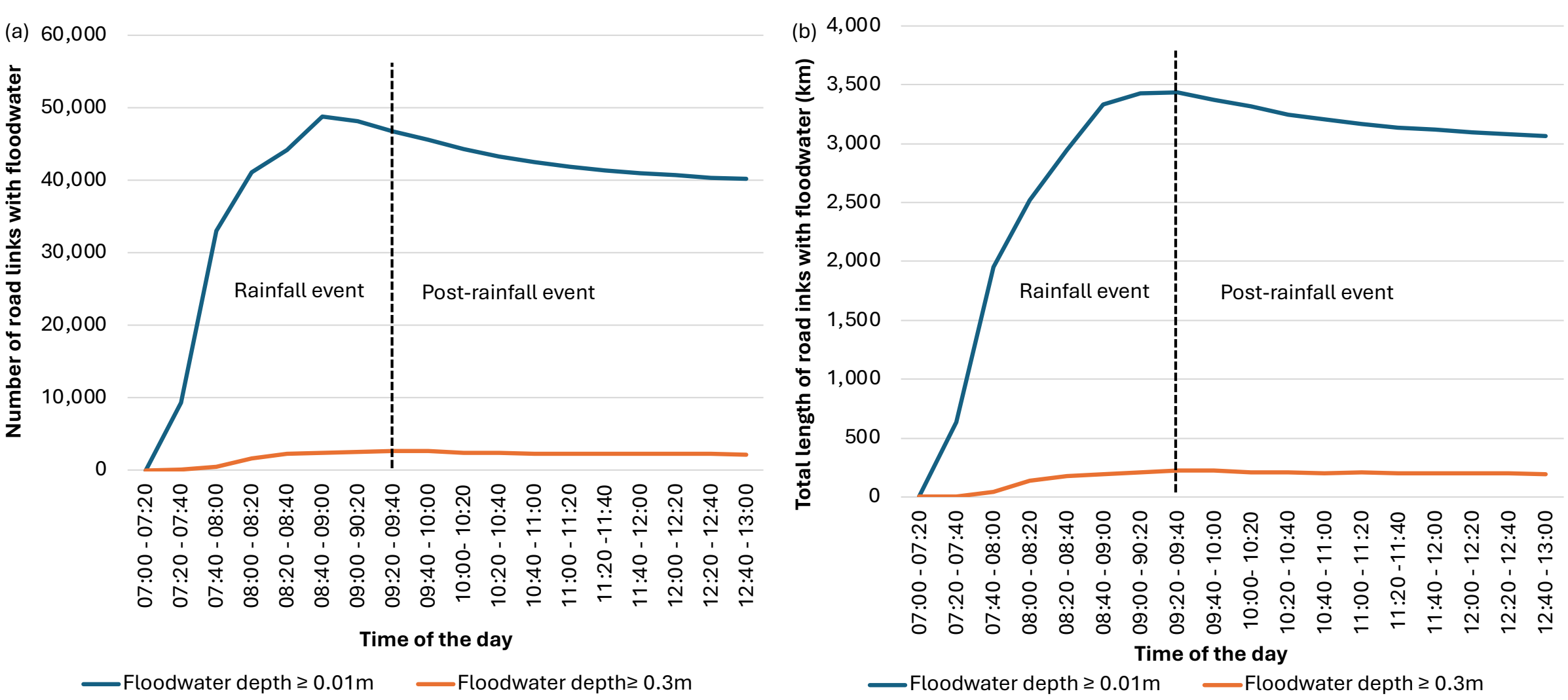

**Fig. 9** Reduction in road network availability over time

Following those peaks, steady but slow decreases are observed in both the number and total length of road links affected by flooding. Six hours after the onset of the rainfall, 40,145 and 3,067 road links are still affected with at least 0.01 m and 0.3 m floodwater depths, respectively, indicating respective decreases of around 18% and 17% (Fig. 9(a)). The corresponding road lengths drop to 3,067 km and 195 km, indicating 11% and 14% reduction from their respective peaks (Fig. 9(b)).

### *4.2 Operational performance loss*

Journey cancellation and delay are used to measure operational performance loss caused by rainfall and resulting flooding. Regarding journey cancellation, Fig. 10 summarises the total number and percentage of incomplete journeys by trip purpose across scenarios S1 and S2. In the worst-case scenario S1, 126,656 journeys are cancelled, which is 21% of the total simulated trips. Journeys returning home accounted for the largest share of cancellation, making up 57% of all cancelled trips and 27.2% of the total simulated home journeys. This is because, in S1, agents located in flooded areas are unable to adapt their behaviours (e.g., altering their routes when the original ones become inaccessible) and thus remained stranded, unable to complete their remaining trips for the day. Additionally, 17.0% of all simulated medical journeys, 15.5% of work journeys, and 9.8% of education journeys are unable to complete in S1, indicating potential disruptions to essential travel in the event of heavy rainfall.

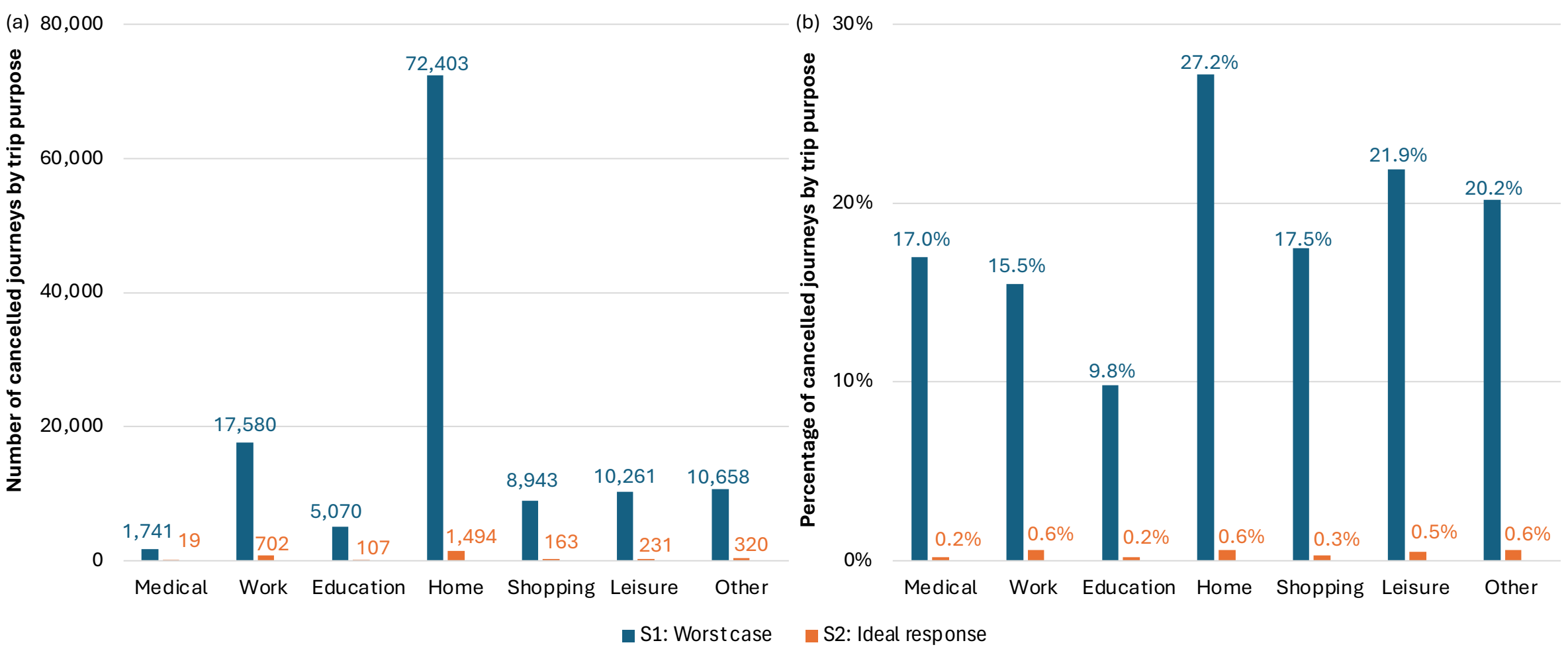

**Fig. 10** Number and percentage of cancelled journeys by trip purpose in scenarios S1 and S2

The number of incomplete journeys is significantly reduced to 3,036 trips in scenario S2 (0.5% of total trips), where agents are allowed to explore alternative activity plans to avoid affected road links. Cancelled journeys account for only a small proportion within each travel purpose group, highlighting the effectiveness of behavioural adaptation during heavy rainfall emergency and the importance of incorporating such behaviour changes into transport modelling. In scenario S3 where early warning is provided, all journeys are completed as a result of reduced travel demand on the road network, indicating the critical role of proactive information dissemination in mitigating disruption and enhancing road network resilience. In comparison with the baseline scenario, 20%, 49%, and 14% of all simulated journeys experience delays in scenarios S1, S2, and S3, respectively (Fig. 11(a)). The lower proportion in S1 is due to the large number of cancelled journeys. The higher proportion in S2 is driven by rerouting effects, whereby rerouted agents impose additional delays on other journeys. While the number of delayed trips during the morning peak is lower in S2 and S3 than in S1, opposite results are observed during the afternoon peak, which is likely due to the journeys cancelled in S1 being partially (in S2) or fully (in S3) completed in the other scenarios. The total delay of all trips reaches 147,408, 41,471, and 10,992 hours in S1, S2, and S3, respectively, with corresponding average delays around 37.1 mins, 8.6 mins, and 7.8 mins (Fig. 11(b)). The average morning peak delay in S1 is approximately 57.2 mins in S1, compared to 8.7 mins in S2 and 13.3 mins in S3, further demonstrating the effectiveness of behavioural adaptation. Finally, the average delay by trip purpose across three scenarios is shown in Fig. 11(c), with only minor variations observed between trip purposes in S2 and S3.

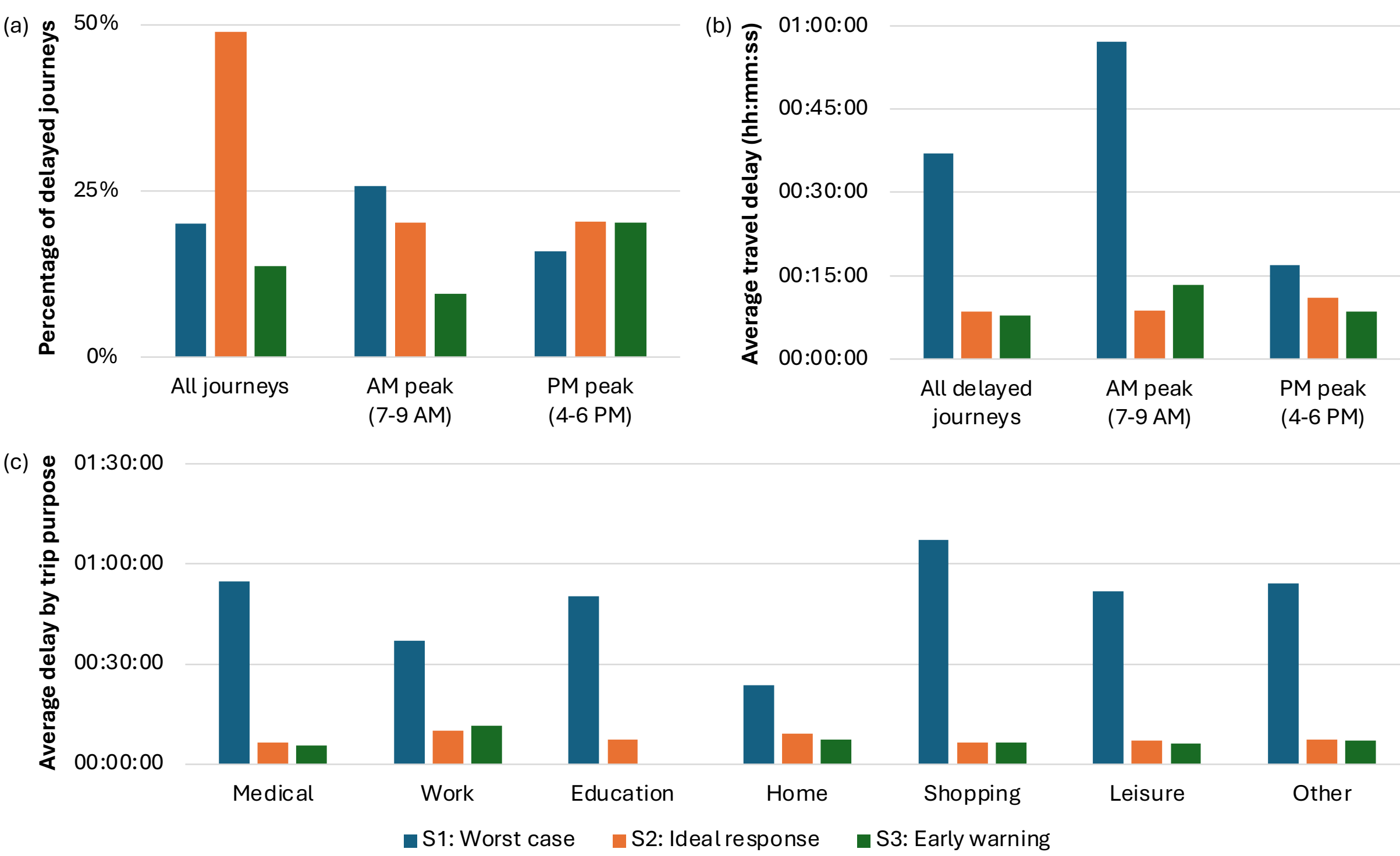

**Fig. 11** Statistics of journey delay in each scenario

### *4.3 Environmental impact*

As a result of the large number of incomplete journeys in S1 and the exclusion of stay-at-home journeys from the simulation in S3, increased emissions from cars are only observed in S2. These include 11,834.0 kgCO2e of GHGs, 22.1 kg of $NO_x$ and 0.3 kg of $PM_{2.5}$, estimated based on a simplified measure using increased travel distance and emission factors. It is worth noting that emission factors are speed-dependent (Chen et al., 2022; J. Wang et al., 2025), which is not considered in this case study but represents an aspect for improvement in future agent-based transport modelling.

### *4.4 Social impact*

To identify potential equity issues induced by disruptions, the percentages of individuals experiencing journey cancellation and delay across demographic groups are analysed, considering age, gender, income, and employment status. Based on the results presented in Fig. 12 and Fig. 13, there is no clear evidence showing that vulnerable groups, such as children, the elderly, women, low-income individuals, and the unemployed, would experience disproportionate impacts on their journeys compared to other less vulnerable populations. For instance, journey cancellation is more prevalent among young to middle-aged adults, individuals with moderate to high incomes, and those who are employed (Fig. 12). This pattern likely reflects higher travel demand and thereby greater exposure to disruptions among these demographic groups. No significant gender-based disparity is observed in this regard. A similar pattern is found in the statistics for individuals experiencing delay in S2, which is given as an illustrative example (Fig. 13). Although vulnerable groups do not appear to experience disproportionate impacts, the absolute level of disruption faced by some groups is noticeable and important to address due to their limited capacity to cope. For example, 33% of students and 17% of child students in S1 are unable to complete their journeys via their normal routes. Finally, a comparison of the statistics for S1 and S2 in Fig. 12 clearly shows the substantial role of behavioural adaptation in mitigating these social impacts.

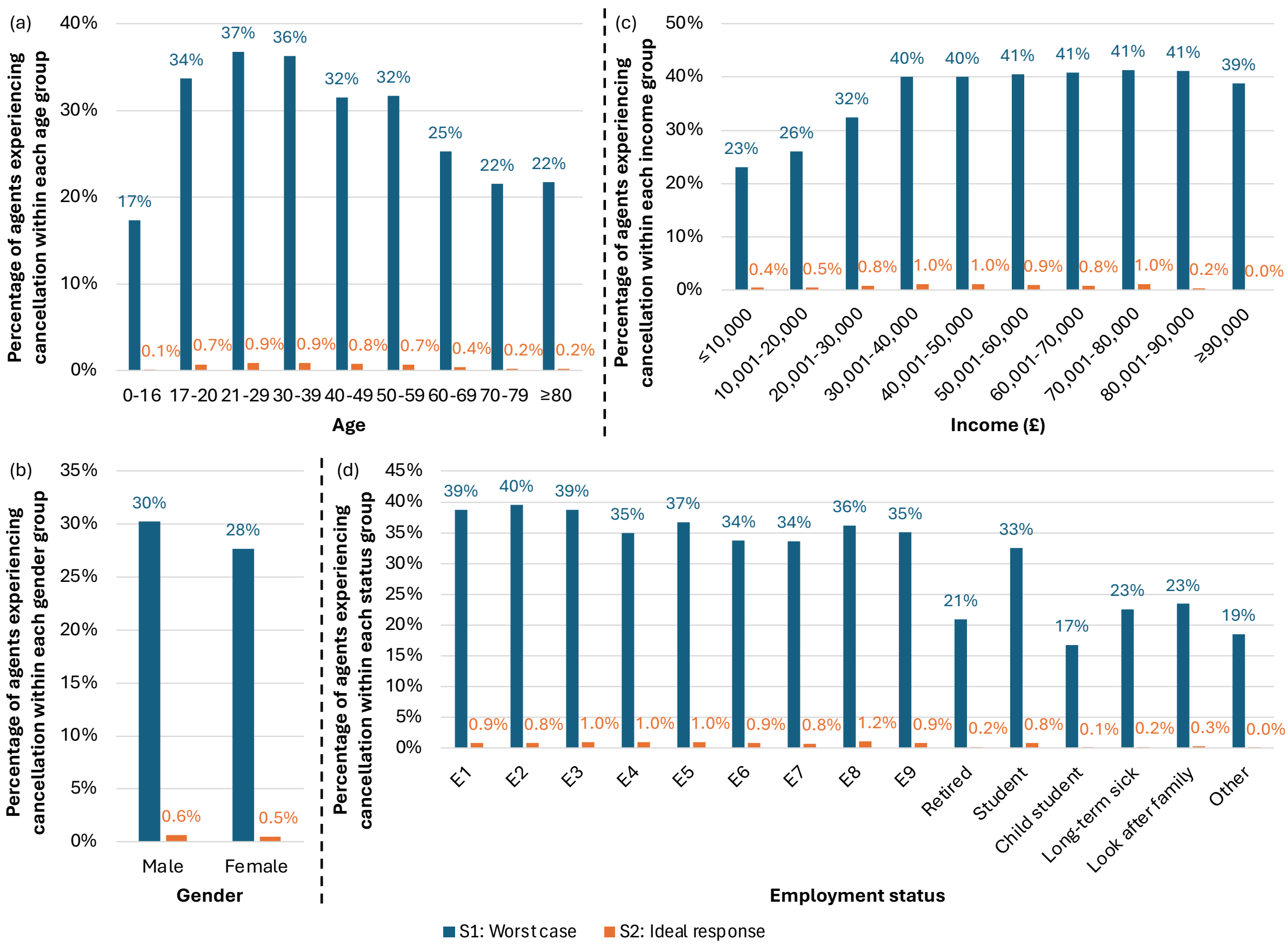

Note: "E" in Figure 10(d) refers to agents being employed. E1: Managers, directors and senior officials. E2: Professional occupations. E3: Associate professional and technical occupations. E4: Administrative and secretarial occupations. E5: Skilled trades occupations. E6: Caring, leisure and other service occupations. E7: Sales and customer service occupations. E8: Process, plant and machine operatives. E9: Elementary occupations.

**Fig. 12** Percentage of agents experiencing journey cancellation by demographic group in S1 and S2

### *4.5 Economic impact*

Based on the unit cost values introduced earlier in Section 3.4, the total cost of journey delay is estimated at approximately £589,275, £318,004, and £87,014 in S1, S2, and S3, respectively. The cost of journey cancellation amounts to £5,649,000 in S1 and £150,340 in S2. These figures suggest that the broader indirect impact of rainfall-induced disruptions could be substantial, underscoring the importance of incorporating adaptive travel behaviour and proactive communication strategies to mitigate cascading societal and economic disruptions during extreme weather events. Regarding societal costs, the impacts are relatively minor in this case. The cost of increased GHG emission is £3,353 in S2, while the cost of increased air pollutant emission amounts to £304. These results indicate that the societal costs represent only a small fraction of the total impact within the scope of this 24-hour simulation. However, this aspect would become more significant in the context of prolonged disruptions—such as an extended recovery period following the collapse and reconstruction of a critical bridge—or when planning for future land use and climate scenarios.

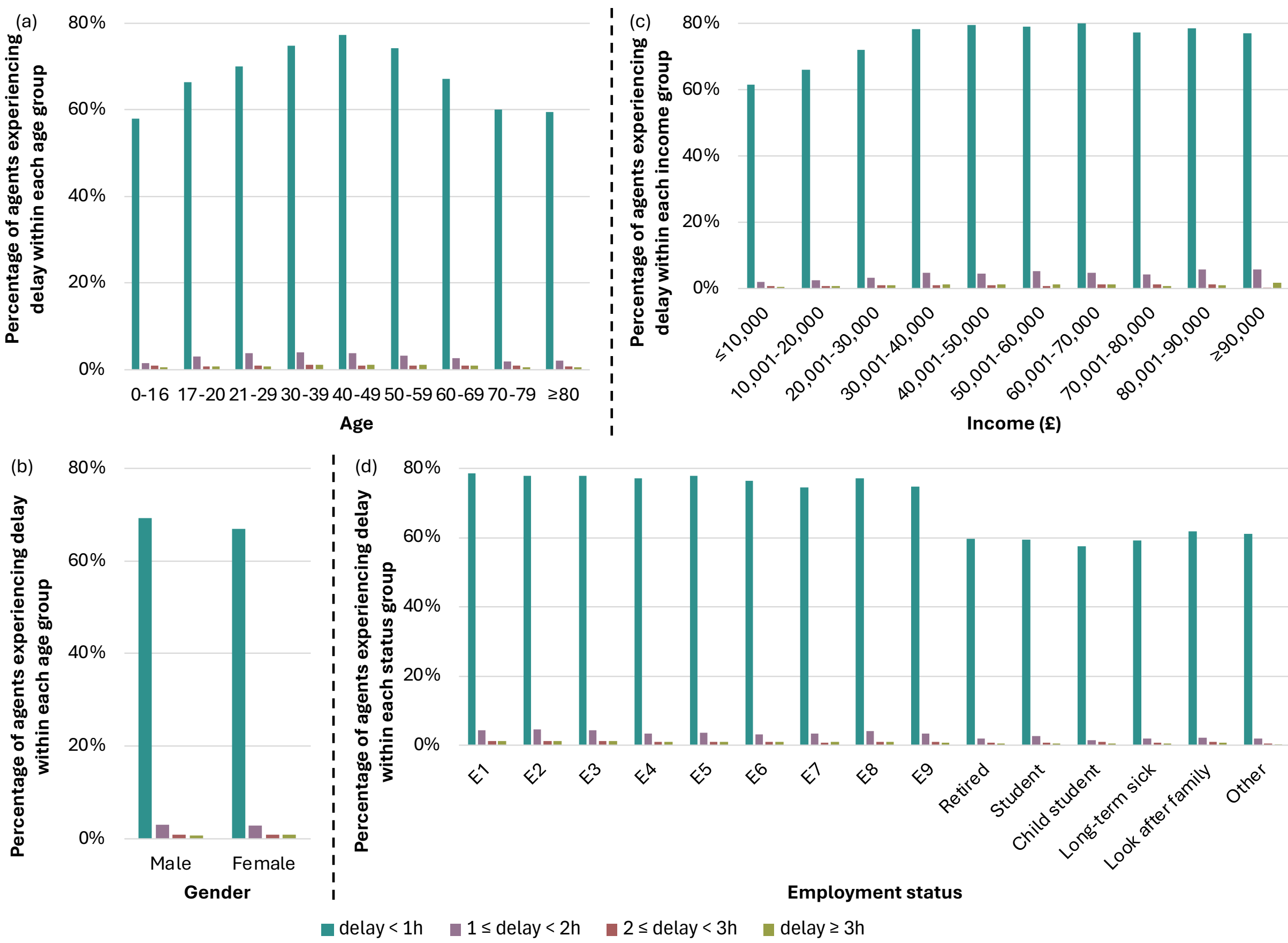

Note: "E" in Figure 10(d) refers to agents being employed. E1: Managers, directors and senior officials. E2: Professional occupations. E3: Associate professional and technical occupations. E4: Administrative and secretarial occupations. E5: Skilled trades occupations. E6: Caring, leisure and other service occupations. E7: Sales and customer service occupations. E8: Process, plant and machine operatives. E9: Elementary occupations.

**Fig. 13** Percentage of individuals experiencing journey delay by demographic group in S2

## 5. Discussion

### *5.1 Effectiveness of "soft adaptation"*

The noticeable reductions in journey delay and cancellation across trip purposes and demographic groups in S2 and S3 demonstrate the pivotal role of behavioural adaptation and early warning in alleviating the short-term disruptive impacts of extreme rainfall on road passenger transport. These adaptative responses have great potential to reduce individual flood risk exposure and limit ripple societal and economic consequences of physical road disruption, such as increased GHG emissions, reduced access to work and education (S2), and economic losses associated with travel delays. As a "soft adaptation" strategy requiring no physical infrastructure intervention, behavioural adaptation and early warning offer a flexible and cost-effective complement to traditional engineering approaches aimed at enhancing asset or system robustness through construction. The case study results underscore the value of integrating behavioural responses, cross-agency coordination, and real-time information dissemination into resilience planning of road passenger transport in response to acute weather-related disruptions.

### *5.2 Selection of behavioural scenarios*

The proposed agent-based integrated modelling framework provides a valuable methodology for assessing the impacts of rainfall-induced transport disruptions at the individual journey level, while capturing how these impacts vary across demographic groups under different behavioural scenarios. As with all models that rely on assumptions to represent complex real-world systems, the three selected behavioural scenarios are not intended to capture the full complexity of human behaviour in practice. Instead, they provide stylised representations of plausible adaptive responses, enabling the potential effects of behavioural adaptation to be explored in a transparent and interpretable way.

To be specific, agents are not allowed to adapt their behaviours in the worst-case scenario S1, whereas in the ideal-response scenario S2, they can adapt behaviours even before the extreme event has started or affected their routes. While these two scenarios provide valuable boundaries for assessing potential impacts, neither fully reflects how individuals might behave under reality, where responses are likely to be delayed, uncertain, and influenced by incomplete or evolving information. Despite this, these scenarios are useful for policy makers to identify the most affected zones in terms of mobility (S1) and to prioritise routes based on the event severity at different timeframes and the number of users (S2). The early-warning scenario S3 keeps specific populations at home based on their socio-demographic characteristics. Although altering the selected population groups and their stay-at-home ratios would yield different outcomes, S3 illustrates how the effectiveness of early-warning strategies can be tested for specific population groups, while also showing how traffic conditions evolve during an extreme rainfall event when selected agents stay at home. In practice, such scenario design could be further supported by behavioural surveys on how different population groups perceive, receive, and respond to extreme-weather warnings, thereby improving the realism of stay-at-home assumptions and increasing confidence in early-warning-based resilience planning.

### *5.3 Extension of rainfall scenarios*

In this case study, a 3-hour, 1-in-30-year intensive rainfall event is used as an illustrative scenario to demonstrate the proposed modelling framework, while the framework itself can be applied to other rainfall scenarios with different assumptions and parameter settings. Rainfall parameters, including intensity, duration, geographical location, and time of occurrence, can affect the resulting impact assessment (Liu et al., 2025a). Future work should consider simulating the same scenarios with variations to understand their potential range of outcomes and their implications for policymakers in decision-making. This could be formulated in a way similar to an ensemble weather modelling approach, with small perturbations applied to the input rainfall data or model parameters. For example, spatial and temporal shifts can be applied to the rainfall data to assess disruption impacts vary under alternative rainfall scenarios. A sufficient variety of rainfall cases with diverse patterns (e.g. linearly organised convective storms moving different directions, or wider slow-moving mesoscale convective systems) could be used to support broader conclusions. Additionally, a limitation of the case study is that only a three-hour recession period is simulated, after which floodwater depth is assumed to remain constant. Future work should extend the modelling to capture the full recession process until complete dissipation.

### *5.4 General applicability of POESE metrics for extreme weather events*

Although derived from the analysis of impact chains associated with rainfall-induced flood events, the proposed multidimensional POESE performance metrics are generally applicable to analyse the disruption impacts of other extreme weather events. Taking heatwave as an example, from the physical road asset perspective, high and sustained temperatures can cause deterioration of bituminous road surfaces through flushing or bleeding (PH1-3), as well as pavement softening that increases the risk of surface deformation and tyre bursts (Matini et al., 2022; McEvoy et al., 2012; Radford et al., 2022; Ukkusuri et al., 2024). This will not only lead to increased maintenance, repair, and even reconstruction costs (EC5-6) but also increase road accident risks as a safety concern (Markolf et al., 2019; Nazif-Munoz et al., 2025) (SO8-10 and EC9). Moreover, on the one hand, extreme heat will reduce the feasibility of active travel (Chen et al., 2025; Li et al., 2024; Wang et al., 2024) and result in an increase in modal shift towards cars (Markolf et al., 2019), which may give rise to traffic congestion and thereby delays (OP1-4) and associated environmental and economic impacts (EN1-3, EC1, and EC7-8). On the other hand, vehicle overheating during heatwaves can yield reduced mobility (Jaroszweski et al., 2014), leading to cancelled journeys (OP5-6 and EC2-3) and thereby a reduction in traffic demand. These opposing effects introduce uncertainty regarding net congestion outcomes during heatwaves, underscoring the need for empirical evidence from historical events or surveys to support modelling assumptions. This example demonstrates how the proposed POESE performance metrics can be flexibly applied to capture complex and potentially competing disruption mechanisms under other extreme weather events beyond rainfall.

## 6 Conclusions

This study proposes a set of multidimensional performance metrics, referred to as POESE, for quantifying flood-induced disruption impacts in road transport systems following heavy rainfall events. By explicitly capturing physical, operational, environmental, social, and economic impacts, the POESE metrics move beyond existing indicators that monitor annual system performance under typical weather conditions or solely consider network connectivity, providing a structured framework of assessing road transport resilience to extreme weather events.

To support the assessment of the proposed metrics, an integrated modelling framework that combines rainfall modelling, surface water flood modelling, and agent-based transport modelling has been developed. This approach enables the simulation of both asset-level disruptions and individual-level behavioural responses, allowing disruption impacts to be evaluated at multiple spatial and temporal scales. Behavioural scenario-based analyses further illustrate how different behavioural responses, including the worst-case, ideal response, and early warning scenarios, can influence disruption consequences.

This methodology is demonstrated through a case study of the Tyne and Wear region in North East England, using a 1-in-30-year rainfall event. Results indicate great potential of behavioural adaptation in enhancing road flood resilience by reducing journey delays, trip cancellations, and economic losses across trip purposes and demographic groups. The integrated modelling framework, together with POESE performance metrics, is readily adaptable for stress-testing various rainfall scenarios and evaluating the effectiveness of resilience interventions for minimising disruption impacts. Furthermore, while the proposed metrics are initially derived from rainfall-induced flood events, they are conceptually applicable to other extreme weather hazards, such as heatwaves.

Moving forward, future work will focus on enhancing the representation of behavioural responses by incorporating delayed and uncertain responses informed by transport user surveys, real-time information access, such as radio and social media. It will also explore the effects of different rainfall timings, intensities, locations, and durations, and extend social impact analysis to account for intersecting demographic characteristics (e.g., low-income women with caregiving responsibilities) and residential location. Simulations with different variations of these components will enable a more comprehensive assessment of disruption impacts under increasingly heavy rainfall events, which could support transport operators in decision-making.

**CRediT authorship contribution statement**

**Wei Bi:** Conceptualization, Methodology, Formal analysis, Writing – original draft, Writing – review & editing. **David Alvarez Castro:** Conceptualization, Data curation, Methodology, Formal analysis, Writing – original draft, Writing – review & editing. **Yimeng Liu:** Methodology, Visualization, Writing – original draft. **Abdullah Kahraman:** Methodology, Visualization, Writing – original draft. **Alistair Ford:** Methodology, Writing – review & editing, Supervision, Funding acquisition. **Roberto Palacin:** Writing – review & editing, Funding acquisition. **Kristen MacAskill:** Writing – review & editing, Supervision, Funding acquisition.

**Declaration of competing interest**

The authors declare that they have no known competing financial interests or personal relationships that could have appeared to influence the work reported in this paper.

**Data availability**

Data will be made available on request.

**Acknowledgements**

The authors acknowledge funding from the UK Engineering & Physical Sciences Research Council (EPSRC) and UK Department for Transport joint grant for the Research Hub for Decarbonised, Adaptable, and Resilient Transport Infrastructures (DARe) (EP/Y024257/1).

**References**

Allalou, R., Hebib, R., Bersi, M., Lefkir, A., 2026. BIM-GIS integrated hydrodynamic modeling for road flood risk assessment: A case study in Algeria's Chiffa Wadi. Reliability Engineering & System Safety 267, 111906. https://doi.org/10.1016/j.ress.2025.111906

Álvarez Castro, D., 2024. An Agent-Based Model framework to simulate active travel-focused transport policy scenarios (PhD Thesis). Newcastle University.

Alvarez Castro, D., Ford, A., James, P., Palacín, R., Ziemke, D., 2024. A MATSim model methodology to generate cycling-focused transport scenarios in England. Journal of Urban Mobility 5, 100078. https://doi.org/10.1016/j.urbmob.2024.100078

Arabi, M., Hyun, K.K., Mattingly, S.P., 2023. Identifying the Most Critical Evacuation Links Based on Road User Vulnerabilities. Natural Hazards Review 24, 04023006. https://doi.org/10.1061/NHREFO.NHENG-1667

Azargoshasbi, R., Esfeh, M.A., Kattan, L., 2026. Urban road network resilience assessment framework: Integrating spatiotemporal analysis with the resilience triangle and temporal performance indicators. Reliability Engineering & System Safety 266, 111742. https://doi.org/10.1016/j.ress.2025.111742

Bhatta, N., Tanim, S.H., Murray-Tuite, P., 2024. Dynamics of Link Importance through Normal Conditions, Flood Response, and Recovery. Sustainability 16, 819. https://doi.org/10.3390/su16020819

Bi, W., Linkov, I., MacAskill, K., 2025. Supporting urban sustainability through resilient rail transit systems. Cities 167, 106373. https://doi.org/10.1016/j.cities.2025.106373

Bi, W., MacAskill, K., Schooling, J., 2023. Old wine in new bottles? Understanding infrastructure resilience: Foundations, assessment, and limitations. Transportation Research Part D: Transport and Environment 120, 103793. https://doi.org/10.1016/j.trd.2023.103793

Bi, W., Schooling, J., MacAskill, K., 2024. Assessing flood resilience of urban rail transit systems: Complex network modelling and stress testing in a case study of London. Transportation Research Part D: Transport and Environment 134, 104263. https://doi.org/10.1016/j.trd.2024.104263

Bruneau, M., Chang, S.E., Eguchi, R.T., Lee, G.C., O'Rourke, T.D., Reinhorn, A.M., Shinozuka, M., Tierney, K., Wallace, W.A., von Winterfeldt, D., 2003. A Framework to Quantitatively Assess and Enhance the Seismic Resilience of Communities. Earthquake Spectra 19, 733–752. https://doi.org/10.1193/1.1623497

Casali, Y., Heinimann, H.R., 2019. A topological characterization of flooding impacts on the Zurich road network. PLoS one 14, e0220338. https://doi.org/10.1371/journal.pone.0220338

Charypar, D., Horni, A., Kickhöfer, B., Nagel, K., 2016. A closer look at scoring, in: The Multi-Agent Transport Simulation MATSim. Ubiquity Press, pp. 23–34.

Charypar, D., Nagel, K., 2005. Generating complete all-day activity plans with genetic algorithms. Transportation 32, 369–397. https://doi.org/10.1007/s11116-004-8287-y

Chen, X., Jiang, L., Xia, Y., Wang, L., Ye, J., Hou, T., Zhang, Yibo, Li, M., Li, Z., Song, Z., Li, J., Jiang, Y., Li, P., Zhang, X., Zhang, Yang, Rosenfeld, D., Seinfeld, J.H., Yu, S., 2022. Quantifying on-road vehicle emissions during traffic congestion using updated emission factors of light-duty gasoline vehicles and real-world traffic monitoring big data. Science of The Total Environment 847, 157581. https://doi.org/10.1016/j.scitotenv.2022.157581

Chen, Z., Qiao, R., Li, S., Zhou, S., Zhang, X., Wu, Z., Wu, T., 2025. Heat and mobility: Machine learning perspectives on bike-sharing resilience in Shanghai. Transportation Research Part D: Transport and Environment 142, 104692. https://doi.org/10.1016/j.trd.2025.104692

Coleman, N., Li, X., Comes, T., Mostafavi, A., 2024. Weaving equity into infrastructure resilience research: a decadal review and future directions. npj Nat. Hazards 1, 25. https://doi.org/10.1038/s44304-024-00022-x

Contreras-Jara, M., Echaveguren, T., Allen, E., Chamorro, A., Vargas-Baecheler, J., 2025. Indirect costs of floods: a case study of highways road users. EGUsphere 1–28. https://doi.org/10.5194/egusphere-2025-4016

Department for Energy Security and Net Zero, 2025. Greenhouse gas reporting: conversion factors 2025. https://www.gov.uk/government/publications/greenhouse-gas-reporting-conversion-factors-2025 (accessed 30 January 2026).

Department for Transport, 2026. National Travel Survey. https://www.gov.uk/government/collections/national-travel-survey-statistics (accessed 22 February 2026).

Department for Transport, 2024. Transport Analysis Guidance. https://www.gov.uk/guidance/transport-analysis-guidance-tag (accessed 19 February 2025).
Department for Transport, 2020. Road investment strategy 2: 2020-2025. https://www.gov.uk/government/publications/road-investment-strategy-2-ris2-2020-to-2025 (accessed 16 December 2024).
Department for Transport, 2014. Transport Resilience Review: A Review of the Resilience of the Transport Network to Extreme Weather Events. https://www.gov.uk/government/publications/transport-resilience-review-recommendations (accessed 5 December 2021).
Digimap, 2020. Ordnance Survey Roam. https://digimap.edina.ac.uk/shibboleth-ds/DS?entityID=https%3A%2F%2Fgeoshibb.edina.ac.uk%2Fshibboleth&return=https%3A%2F%2Fdigimap.edina.ac.uk%2FShibboleth.sso%2FDS%3FSAMLDS%3D1%26target%3Dcookie%253A1769699365_1ffe (accessed 16 July 2025).
Dong, S., Gao, X., Mostafavi, A., Gao, J., Gangwal, U., 2023. Characterizing resilience of flood-disrupted dynamic transportation network through the lens of link reliability and stability. Reliability Engineering & System Safety 232, 109071. https://doi.org/10.1016/j.ress.2022.109071
Du, Q., Zong, X., Li, Y., Guo, X., Ye, Z., Li, S., Bai, L., 2024. Resilience optimization of bus-metro double-layer network against extreme weather events. Transportation Research Part D: Transport and Environment 135, 104378. https://doi.org/10.1016/j.trd.2024.104378
Franco, P., Johnston, R., McCormick, E., 2020. Demand responsive transport: Generation of activity patterns from mobile phone network data to support the operation of new mobility services. Transportation Research Part A: Policy and Practice, Developments in Mobility as a Service (MaaS) and Intelligent Mobility 131, 244–266. https://doi.org/10.1016/j.tra.2019.09.038
Gangwal, U., Siders, A.R., Horney, J., Michael, H.A., Dong, S., 2023. Critical facility accessibility and road criticality assessment considering flood-induced partial failure. Sustainable and Resilient Infrastructure 8, 337–355. https://doi.org/10.1080/23789689.2022.2149184
Glenis, V., Kutija, V., Kilsby, C.G., 2018. A fully hydrodynamic urban flood modelling system representing buildings, green space and interventions. Environmental Modelling & Software 109, 272–292. https://doi.org/10.1016/j.envsoft.2018.07.018
Greater Manchester Transport Committee, 2023. Bus Performance Report. https://democracy.greatermanchester-ca.gov.uk/documents/s25182/Item%206%20-%20GMTC%20Bus%2020230310%20Bus%20Performance%20Report.pdf (accessed 26 December 2024).
Haritha, P.C., Anjaneyulu, M.V.L.R., 2024. Comparison of topological functionality-based resilience metrics using link criticality. Reliability Engineering & System Safety 243, 109881. https://doi.org/10.1016/j.ress.2023.109881
He, K., Carhart, N., Pregnolato, M., De Risi, R., 2026. Flood-induced traffic congestion and accessibility loss for urban road networks using agent-based simulation: The case study of Bristol, UK. International Journal of Disaster Risk Reduction 136, 106053. https://doi.org/10.1016/j.ijdrr.2026.106053
He, K., Pregnolato, M., Carhart, N., Neal, J., De Risi, R., 2024. Functionality assessment of road network combining flood roadworthiness and graph topology. Transportation Research Part D: Transport and Environment 135, 104354. https://doi.org/10.1016/j.trd.2024.104354

HM Government, 2019. Environmental Reporting Guidelines: Including streamlined energy and carbon reporting guidance. https://assets.publishing.service.gov.uk/media/67161e8696def6d27a4c9ab3/environmental-reporting-guidance-secr-march-2019.pdf (accessed 26 February 2025).
Horni, A., Nagel, K., Axhausen, K.W., 2016. The Multi-Agent Transport Simulation MATSim. London.
Huang, J., Cui, Y., Zhang, L., Tong, W., Shi, Y., Liu, Z., 2022. An Overview of Agent-Based Models for Transport Simulation and Analysis. Journal of Advanced Transportation 2022, 1252534. https://doi.org/10.1155/2022/1252534
Iliadis, C., Galiatsatou, P., Glenis, V., Prinos, P., Kilsby, C., 2023. Urban Flood Modelling under Extreme Rainfall Conditions for Building-Level Flood Exposure Analysis. Hydrology 10, 172. https://doi.org/10.3390/hydrology10080172
Jaroszweski, D., Hooper, E., Chapman, L., 2014. The impact of climate change on urban transport resilience in a changing world. Progress in Physical Geography: Earth and Environment 38, 448–463. https://doi.org/10.1177/0309133314538741
Johnston, R.A., 2004. The urban transportation planning process.
Kendon, E., Short, C., Pope, J., Chan, S., Wilkinson, J., Tucker, S., Bett, P., Harris, G., 2021. Update to UKCP Local (2.2km) projections. https://www.metoffice.gov.uk/pub/data/weather/uk/ukcp18/science-reports/ukcp18_local_update_report_2021.pdf (accessed 1 June 2025).
Kendon, E.J., Roberts, N.M., Fosser, G., Martin, G.M., Lock, A.P., Murphy, J.M., Senior, C.A., Tucker, S.O., 2020. Greater Future U.K. Winter Precipitation Increase in New Convection-Permitting Scenarios. Journal of Climate 33, 7303–7318. https://doi.org/10.1175/JCLI-D-20-0089.1
Kim, K., Kaviari, F., Pant, P., Yamashita, E., 2022. An agent-based model of short-notice tsunami evacuation in Waikiki, Hawaii. Transportation Research Part D: Transport and Environment 105, 103239. https://doi.org/10.1016/j.trd.2022.103239
Koks, E.E., Rozenberg, J., Zorn, C., Tariverdi, M., Vousdoukas, M., Fraser, S.A., Hall, J.W., Hallegatte, S., 2019. A global multi-hazard risk analysis of road and railway infrastructure assets. Nat Commun 10, 2677. https://doi.org/10.1038/s41467-019-10442-3
Li, C., Chen, G., Wang, S., 2024. Urban mobility resilience under heat extremes: Evidence from bike-sharing travel in New York. Travel Behaviour and Society 37, 100821. https://doi.org/10.1016/j.tbs.2024.100821
Li, C., Wang, W., Solé-Ribalta, A., Borge-Holthoefer, J., Jia, B., Liu, Z., Yu, B., Gao, Z., Gao, J., 2025. Adaptive capacity for multimodal transport network resilience to extreme floods. Nat Sustain 1–12. https://doi.org/10.1038/s41893-025-01575-z
Li, L., Li, B., Zhong, L., 2026. Resilience assessment of urban mobility flow networks from different scales: A case study in shenzhen. Reliability Engineering & System Safety 274, 112374. https://doi.org/10.1016/j.ress.2026.112374
Li, M., Song, Y.-R., Song, B., Jiang, G.-P., 2026. Resilience assessment and enhancement of urban transportation interdependent network under cascading failure. Reliability Engineering & System Safety 273, 112302. https://doi.org/10.1016/j.ress.2026.112302
Li, Y., Pant, R., Russell, T., Fred Thomas, Hall, J.W., Oldham, P., Lamb, R., Young, P.J., 2026. Stress-testing road network resilience using counterfactual flood events (1953–2024) in Great Britain.

Transportation Research Part D: Transport and Environment 155, 105292. https://doi.org/10.1016/j.trd.2026.105292
Linkov, I., Trump, B.D., Trump, J., Pescaroli, G., Hynes, W., Mavrodieva, A., Panda, A., 2022. Resilience stress testing for critical infrastructure. International Journal of Disaster Risk Reduction 82, 103323. https://doi.org/10.1016/j.ijdrr.2022.103323
Litman, T., 2013. Transportation and Public Health. Annual Review of Public Health 34, 217–233. https://doi.org/10.1146/annurev-publhealth-031912-114502
Liu, J., Xu, Z., Wan, L., MacAskill, K., 2026a. Beyond waterlogging: Evaluating the impact of extreme rainfall on the road network. Reliability Engineering & System Safety 273, 112308. https://doi.org/10.1016/j.ress.2026.112308
Liu, J., Xu, Z., Wan, L., MacAskill, K., 2026b. Road Transport Resilience under Extreme Rainfall: Integrating Multiple Impact Factors and Delay Propagation. International Journal of Disaster Risk Reduction 106024. https://doi.org/10.1016/j.ijdrr.2026.106024
Liu, K., Wang, Q., Wang, M., Koks, E.E., 2023. Global transportation infrastructure exposure to the change of precipitation in a warmer world. Nat Commun 14, 2541. https://doi.org/10.1038/s41467-023-38203-3
Liu, Y., Ford, A., Dawson, R., Pyatkova, K., Tang, J., Wan, Y., Yang, S., 2025a. Flood impact on urban transport networks considering the flooding propagation. Journal of Environmental Management 395, 127972. https://doi.org/10.1016/j.jenvman.2025.127972
Liu, Y., Ford, A., Dawson, R., Pyatkova, K., Tang, J., Wan, Y., Yang, S., 2025b. Flood impact on urban transport networks considering the flooding propagation. Journal of Environmental Management 395, 127972. https://doi.org/10.1016/j.jenvman.2025.127972
Lu, Q.-L., Sun, W., Dai, J., Schmöcker, J.-D., Antoniou, C., 2024. Traffic resilience quantification based on macroscopic fundamental diagrams and analysis using topological attributes. Reliability Engineering & System Safety 247, 110095. https://doi.org/10.1016/j.ress.2024.110095
Ma, R., Dong, H., Han, Q., Du, X., 2025. Resilience modeling of transportation infrastructure and network based on the semi-Markov process considering resource dependency. Reliability Engineering & System Safety 261, 111159. https://doi.org/10.1016/j.ress.2025.111159
Manley, E., Filomena, G., Mavros, P., 2021. A spatial model of cognitive distance in cities. International Journal of Geographical Information Science 35, 2316–2338. https://doi.org/10.1080/13658816.2021.1887488
Markolf, S.A., Hoehne, C., Fraser, A., Chester, M.V., Underwood, B.S., 2019. Transportation resilience to climate change and extreme weather events – Beyond risk and robustness. Transport policy 74, 174–186.
Mathew, S., Pulugurtha, S.S., 2022. Quantifying the Effect of Rainfall and Visibility Conditions on Road Traffic Travel Time Reliability. Weather, Climate, and Society 14, 507–519. https://doi.org/10.1175/WCAS-D-21-0053.1
Matini, N., Gulzar, S., Underwood, S., Castorena, C., 2022. Evaluation of Structural Performance of Pavements under Extreme Events: Flooding and Heatwave Case Studies. Transportation Research Record 2676, 233–248. https://doi.org/10.1177/03611981221077984
McEvoy, D., Ahmed, I., Mullett, J., 2012. The impact of the 2009 heat wave on Melbourne's critical infrastructure. Local Environment 17, 783–796. https://doi.org/10.1080/13549839.2012.678320
Mersini, T.-M., Ziemke, T., 2026. Scaling MATSim traffic flow properties to properly account for population downsampling. Procedia Computer Science, The 17th International Conference on Ambient Systems, Networks and Technologies Networks (ANT)/ the 9th International Conference on Emerging Data and Industry 4.0 (EDI40) 280, 977–984. https://doi.org/10.1016/j.procs.2026.04.124
Miller, R.L., 2022. Modeling Temporal Accessibility of an Urban Road Network during an Extreme Pluvial Flood Event. Natural Hazards Review 23, 04022032. https://doi.org/10.1061/(ASCE)NH.1527-6996.0000586
Mitropoulos, L., Karolemeas, C., Tsigdinos, S., 2025. Road network accessibility assessment during flood events by using infrastructure facility-based indices. International Journal of Disaster Risk Reduction 122, 105475. https://doi.org/10.1016/j.ijdrr.2025.105475
Mladenovic, M., Trifunovic, A., 2014. The shortcomings of the conventional four step travel demand forecasting process. Journal of Road and Traffic Engineering 60, 5–12.
Moyo Oliveros, M., Nagel, K., 2016. Automatic calibration of agent-based public transit assignment path choice to count data. Transportation Research Part C: Emerging Technologies 64, 58–71. https://doi.org/10.1016/j.trc.2016.01.003
Mukesh, M.S., Katpatal, Y.B., Londhe, D.S., 2022. Measurement of city road network resilience in hazardous flood events. International Journal of Disaster Resilience in the Built Environment 15, 274–288. https://doi.org/10.1108/IJDRBE-11-2021-0155
National Academies of Sciences, Engineering, and Medicine, 2021. Investing in Transportation Resilience: A Framework for Informed Choices. The National Academies Press, Washington, DC. https://doi.org/10.17226/26292 (accessed 1 January 2025).
National Atmospheric Emissions Inventory, 2023. Emission Factors for Transport. http://naei.energysecurity.gov.uk/emission-factors/emission-factors-transport (accessed 14 July 2025).
National Highways, 2024. Performance Monitoring Statements: Year end 2023-24. https://nationalhighways.co.uk/media/hymao3xw/20240719-national-highways-performance-monitoring-statements-year-end-2023-24.pdf (accessed 25 December 2024).
National Highways, 2023. Operational Metrics Manual. https://nationalhighways.co.uk/media/vvsfmz1k/ris2-omm-published-18-july-2023-final.pdf (accessed 3 December 2024).
National Infrastructure Commission, 2024. Developing resilience standards in UK infrastructure. https://nic.org.uk/app/uploads/NIC-Resilience-Standards-Report-Final-190924.pdf (accessed 18 October 2024).
National Infrastructure Commission, 2022. Reducing the risk of surface water flooding. https://nic.org.uk/app/uploads/NIC-Reducing-the-Risk-of-Surface-Water-Flooding-Final-28-Nov-2022.pdf (accessed 20 August 2024).
Nazif-Munoz, J.I., Gilani, V.N.M., Rana, J., Choma, E., Spengler, J.D., Cedeno-Laurent, J.G., 2025. The influence of heatwaves on traffic safety in five cities across Québec with different thermal landscapes. Injury Epidemiology 12, 12. https://doi.org/10.1186/s40621-025-00564-2
Office for National Statistics, 2026a. Nomis - Official Census and Labour Market Statistics. https://www.nomisweb.co.uk/ (accessed 22 February 2026).
Office for National Statistics, 2026b. Home - Office for National Statistics. https://www.ons.gov.uk/ (accessed 22 February 2026).

Office of Rail and Road, 2024a. National Highways Regional Performance Summaries. https://www.orr.gov.uk/sites/default/files/2024-11/national-highways-regional-performance-summaries-2023-24.pdf (accessed 16 December 2024).

Office of Rail and Road, 2024b. Annual assessments of National Highways. https://www.orr.gov.uk/monitoring-and-regulation/roads-monitoring/annual-assessment-national-highways (accessed 26 December 2025).

OpenStreetMap, 2018. Data Extracts. Geofabrik Download Server. https://download.geofabrik.de/ (accessed 27 November 2023).

Panakkal, P., Padgett, J.E., 2024. More eyes on the road: Sensing flooded roads by fusing real-time observations from public data sources. Reliability Engineering & System Safety 251, 110368. https://doi.org/10.1016/j.ress.2024.110368

Pregnolato, M., Ford, A., Wilkinson, S.M., Dawson, R.J., 2017. The impact of flooding on road transport: A depth-disruption function. Transportation research part D: transport and environment 55, 67–81. https://doi.org/10.1016/j.trd.2017.06.020

Pyatkova, K., Chen, A.S., Butler, D., Vojinović, Z., Djordjević, S., 2019. Assessing the knock-on effects of flooding on road transportation. Journal of Environmental Management 244, 48–60. https://doi.org/10.1016/j.jenvman.2019.05.013

Radford, D.A.G., Lawler, T.C., Edwards, B.R., Disher, B.R.W., Maier, H.R., Ostendorf, B., Nairn, J., van Delden, H., Goodsite, M., 2022. A framework for the mitigation and adaptation from heat-related risks to infrastructure. Sustainable Cities and Society 81, 103820. https://doi.org/10.1016/j.scs.2022.103820

Rail Delivery Group, 2023. Rail Industry Data. https://data.atoc.org/ (accessed 22 February 2026).

Railsback, S.F., Grimm, V., 2019. Agent-Based and Individual-Based Modeling: A Practical Introduction, Second Edition. Princeton University Press.

Rieser, M., Nagel, K., Horni, A., 2016. MATSim Data Containers, in: Multi-Agent Transport Simulation MATSim. Ubiquity Press. https://doi.org/10.5334/baw.6

Sattar, K.A., Ishak, I., Affendey, L.S., Mohd Rum, S.N.B., 2024. Road Crash Injury Severity Prediction Using a Graph Neural Network Framework. IEEE Access 12, 37540–37556. https://doi.org/10.1109/ACCESS.2024.3373885

Shang, W.L., Chen, Y.Y., Ochieng, W.Y., 2020. Resilience Analysis of Transport Networks by Combining Variable Message Signs With Agent-Based Day-to-Day Dynamic Learning. IEEE Access 8, 104458–104468. https://doi.org/10.1109/access.2020.2999129

Soriano, A., Gaikwad, S., Stratton-Short, S., Bajpai, A., Imbuye, J., 2022. Inclusive Infrastructure for Climate Action. UNOPS, Copenhagen, Denmark.

Stroeve, S.H., Everdij, M.H.C., 2017. Agent-based modelling and mental simulation for resilience engineering in air transport. Saf. Sci. 93, 29–49. https://doi.org/10.1016/j.ssci.2016.11.003

Sun, L., D'Ayala, D., Fayjaloun, R., Gehl, P., 2021. Agent-based model on resilience-oriented rapid responses of road networks under seismic hazard. Reliability Engineering & System Safety 216, 108030. https://doi.org/10.1016/j.ress.2021.108030

Tang, J., Wu, S., Yang, S., Shi, Y., 2024. Resilience Assessment of Urban Road Transportation in Rainfall. Remote Sensing 16, 3311. https://doi.org/10.3390/rs16173311

Testa, A.C., Furtado, M.N., Alipour, A., 2015. Resilience of coastal transportation networks faced with extreme climatic events. Transportation Research Record 2532, 29–36. https://doi.org/10.3141/2532-04

Theofilatos, A., Yannis, G., 2014. A review of the effect of traffic and weather characteristics on road safety. Accident Analysis & Prevention 72, 244–256. https://doi.org/10.1016/j.aap.2014.06.017

Transport for Greater Manchester, 2024. Greater Manchester Transport Network Performance. https://democracy.greatermanchester-ca.gov.uk/mgConvert2PDF.aspx?ID=31602 (accessed 26 December 2024).

Transport for London, 2024. Customer service and operational performance report: Quarter 1 2024/25. https://tfl.gov.uk/corporate/publications-and-reports/customer-service-op-performance (accessed 25 December 2024).

Transport Scotland, 2016. Road Asset Management Plan for Scottish Trunk Roads. https://www.transport.gov.scot/media/32978/j408891.pdf (accessed 16 December 2024).

Traveline, 2023. Traveline National Dataset. https://www.travelinedata.org.uk/traveline-open-data/traveline-national-dataset/ (accessed 22 February 2026).

Met Office, 2018. UK Climate Projections (UKCP18). https://doi.org/https://www.metoffice.gov.uk/research/approach/collaboration/ukcp (accessed 1 June 2025).

Ukkusuri, S.V., Park, S.U., Mittal, S., Chapman, L., Manoli, G., Santos, A., Jones, N.K.W., Avner, P., Romero, N., 2024. We need to prepare our transport systems for heatwaves — here's how. Nature 632, 253–256. https://doi.org/10.1038/d41586-024-02538-8

Wang, J., Dai, H., Feng, H., Guo, M., Zylianov, V., Feng, Z., Cui, J., 2025. Carbon emission of urban vehicles based on carbon emission factor correlation analysis. Sci Rep 15, 2037. https://doi.org/10.1038/s41598-025-86119-3

Wang, W., Wan, Y., Liu, Q., Liu, H., Li, C., Yang, S., Lai, X., Liu, Y., Yue, Q., 2026. Urban road network resilience under emergencies: a microscopic simulation framework. Transportation Research Part D: Transport and Environment 151, 105158. https://doi.org/10.1016/j.trd.2025.105158

Wang, W., Yang, S., Stanley, H.E., Gao, J., 2019. Local floods induce large-scale abrupt failures of road networks. Nat Commun 10, 2114. https://doi.org/10.1038/s41467-019-10063-w

Wang, X., Xue, R., Lu, M., Wu, J., 2024. Riders Under the Heat: Exploring the Impact of Extreme Heat on the Integration of Bike-Sharing and Public Transportation in Shenzhen, China. ISPRS International Journal of Geo-Information 13, 438. https://doi.org/10.3390/ijgi13120438

Wang, Y., Ye, Z., Jia, X., Liu, H., Zhou, G., Wang, L., 2025. Flood disaster chain deduction based on cascading failures in urban critical infrastructure. Reliability Engineering & System Safety 261, 111160. https://doi.org/10.1016/j.ress.2025.111160

Woodcock, J., Edwards, P., Tonne, C., Armstrong, B.G., Ashiru, O., Banister, D., Beevers, S., Chalabi, Z., Chowdhury, Z., Cohen, A., Franco, O.H., Haines, A., Hickman, R., Lindsay, G., Mittal, I., Mohan, D., Tiwari, G., Woodward, A., Roberts, I., 2009. Public health benefits of strategies to reduce greenhouse-gas emissions: urban land transport. The Lancet 374, 1930–1943. https://doi.org/10.1016/S0140-6736(09)61714-1

Wu, M., Xin, Y., Xiao, L., Zhang, Y., Mei, S., Li, M., 2026. Urban road connectivity assessment and impassable-road restoration sequencing under the impact of flood events. Reliability Engineering & System Safety 274, 112459. https://doi.org/10.1016/j.ress.2026.112459

Xu, W., Zhang, Y., Proverbs, David., Zhong, Z., 2022. Enhancing the resilience of road networks to flooding. International Journal of Building Pathology and Adaptation 42, 213–236. https://doi.org/10.1108/IJBPA-09-2021-0120

Xu, X., Chen, A., Xu, G., Yang, C., Lam, W.H.K., 2021. Enhancing network resilience by adding redundancy to road networks. Transportation Research Part E: Logistics and Transportation Review 154, 102448. https://doi.org/10.1016/j.tre.2021.102448

Yang, Z., Clemente, M.F., Laffréchine, K., Heinzlef, C., Serre, D., Barroca, B., 2022. Resilience of Social-Infrastructural Systems: Functional Interdependencies Analysis. Sustainability 14, 606. https://doi.org/10.3390/su14020606

Zhang, H., Ouyang, M., Xu, M., Sun, W., Hong, L., Li, D., 2026. Travel Time Resilience of National Multimodal Transport Systems under Extreme Events: A Passenger-Oriented Framework. Reliability Engineering & System Safety 112256. https://doi.org/10.1016/j.ress.2026.112256

Zhang, J., Wang, H., Huang, J., Wang, Y., Liu, G., 2023. A study on dynamic simulation and improvement strategies of flood resilience for urban road system. Journal of Environmental Management 344, 118770. https://doi.org/10.1016/j.jenvman.2023.118770